\documentclass[twocolumn,prl]{revtex4-1}
\usepackage{color}
\usepackage{amsmath}
\usepackage{amssymb}
\usepackage{graphicx}
\usepackage[unicode=true,pdfusetitle,
 bookmarks=true,bookmarksnumbered=false,bookmarksopen=false,
 breaklinks=false,pdfborder={0 0 0},pdfborderstyle={},backref=false,colorlinks=true]
 {hyperref}
\hypersetup{colorlinks,linkcolor=red,citecolor=blue}
\usepackage{changes}

\makeatletter

\DeclareFontEncoding{LGR}{}{}
\DeclareRobustCommand{\greektext}{%
  \fontencoding{LGR}\selectfont\def\encodingdefault{LGR}}
\DeclareRobustCommand{\textgreek}[1]{\leavevmode{\greektext #1}}
\ProvideTextCommand{\~}{LGR}[1]{\char126#1}

\usepackage{amsfonts}

\makeatother

\begin{document}

\title{Photorefraction-assisted self-emergence of dissipative Kerr solitons}

\author{Shuai Wan$^{1,2,5}$, Pi-Yu Wang$^{1,2,5}$, Rui Ma$^{3}$, Zheng-Yu
Wang$^{1,2,5}$, Rui Niu$^{1,2,5}$, De-Yong He$^{1,2,5}$, Guang-Can
Guo$^{1,2,5}$, Fang Bo$^{3*}$, Junqiu Liu$^{4,5}$ and Chun-Hua Dong$^{1,2,5*}$}

\affiliation{$^{1}$CAS Key Laboratory of Quantum Information, University of Science
and Technology of China, Hefei, Anhui 230026, People's Republic of
China}

\affiliation{$^{2}$CAS Center For Excellence in Quantum Information and Quantum
Physics, University of Science and Technology
of China, Hefei, Anhui 230088, People's Republic of China}

\affiliation{$^{3}$MOE Key Laboratory of Weak-Light Nonlinear Photonics, TEDA
Applied Physics Institute and School of Physics, Nankai University,
Tianjin 300457, People\textquoteright s Republic of China}

\affiliation{$^{4}$International Quantum Academy, 518048 Shenzhen, China}

\affiliation{$^{5}$Hefei National Laboratory, University of Science and Technology
of China, Hefei, Anhui 230088, People\textquoteright s Republic of
China}
\email{bofang@nankai.edu.cn; chunhua@ustc.edu.cn}

\begin{abstract}
Generated in high-$Q$ optical microresonators, dissipative Kerr soliton microcombs constitute broadband optical frequency combs with chip sizes and repetition rates in the microwave to millimeter-wave range.
For frequency metrology applications such as spectroscopy, optical atomic clocks and frequency synthesizers, octave-spanning soliton microcombs generated in dispersion-optimized microresonator are required, which allow self-referencing for full frequency stabilization.
In addition, field-deployable applications require the generation of such soliton microcombs simple, deterministic, and reproducible.
Here, we demonstrate a novel scheme to generate self-emerging solitons in integrated lithium niobate microresonators.
The single soliton features a broadband spectral bandwidth with dual dispersive waves, allowing $2f$--$3f$ self-referencing.
Via harnessing the photorefractive effect of lithium niobate to significantly extend the soliton existence range, we observe a spontaneous yet deterministic single-soliton formation.
The soliton is immune to external perturbation and can operate continuously over 13 hours without active feedback control.
Finally, via integration with a pre-programed DFB laser, we demonstrate turnkey soliton generation.
 With further improvement of microresonator $Q$ and hybrid integration with chip-scale laser chips, compact soliton microcomb devices with electronic actuation can be created, which can become central elements for future LiDAR, microwave photonics and optical telecommunications.
\end{abstract}
\maketitle

\section{INTRODUCTION}

Dissipative Kerr solitons (DKSs) \cite{Leo:10, Herr:14} are self-organized structures formed in optical microresonators that are driven by continuous-wave (CW) lasers.
Often referred as ``soliton microcombs'' \cite{Kippenberg:18, Pasquazi:18}, they provide a novel route towards broadband, fully coherent, chip-scale frequency combs with small sizes, weight, and power consumption.
Different from commercial fiber-laser-based frequency combs, soliton microcombs feature repetition rates in the microwave to millimeter-wave domain \cite{Yi:15, Suh:18b, Pfeiffer:17, Li:17}.   
Particularly, with advances in fabrication of ultralow-loss integrated waveguides based on many CMOS-compatible materials \cite{Moss:13, Gaeta:19, Kovach:20}, photonic-chip-based soliton microcombs can be manufactured with large volume and low cost.
The merging with integrated photonics has vitalized soliton microcomb technology and accelerated its wider deployment in system-level applications including telecommunication \cite{Marin-Palomo:17, Fujii:22}, astronomical spectrometer calibration \cite{Obrzud:19, Suh:19}, ultrafast ranging \cite{Trocha:18, Suh:18}, neurmorphic photonic computing \cite{Feldmann:21, Xu:21}, microwave generation \cite{Liang:15, Liu:20}, dual-comb spectroscopy \cite{Yang:19, Dutt:18}, quantum key distribution \cite{WangFX:20}, frequency synthesizers \cite{Spencer:18} and optical atomic clocks \cite{Newman:19}.

Critical to applications in frequency metrology \cite{Spencer:18, Newman:19} is the self-referencing of soliton microcombs \cite{Jost:15, DelHaye:16, Brasch:17}.
This requires solitons with spectral bandwidth up to an octave \cite{Brasch:15, Pfeiffer:17, Li:17, Yu:19, LiuX:21, Gong:20} for $f-2f$ or $2f-3f$ self-referencing.
Such a broad bandwidth is typically obtained in microresonators of properly engineered dispersion, where dual-dispersive-wave emission enhances local comb line power and coherently extends the soliton spectrum.
The dispersive waves are generated at frequency positions phase-matched to the CW pump.
The positions can be precisely engineered such that they corresponds to the frequency components for self-referencing.

\begin{figure*}
\centering{}
\includegraphics[clip,width=16cm]{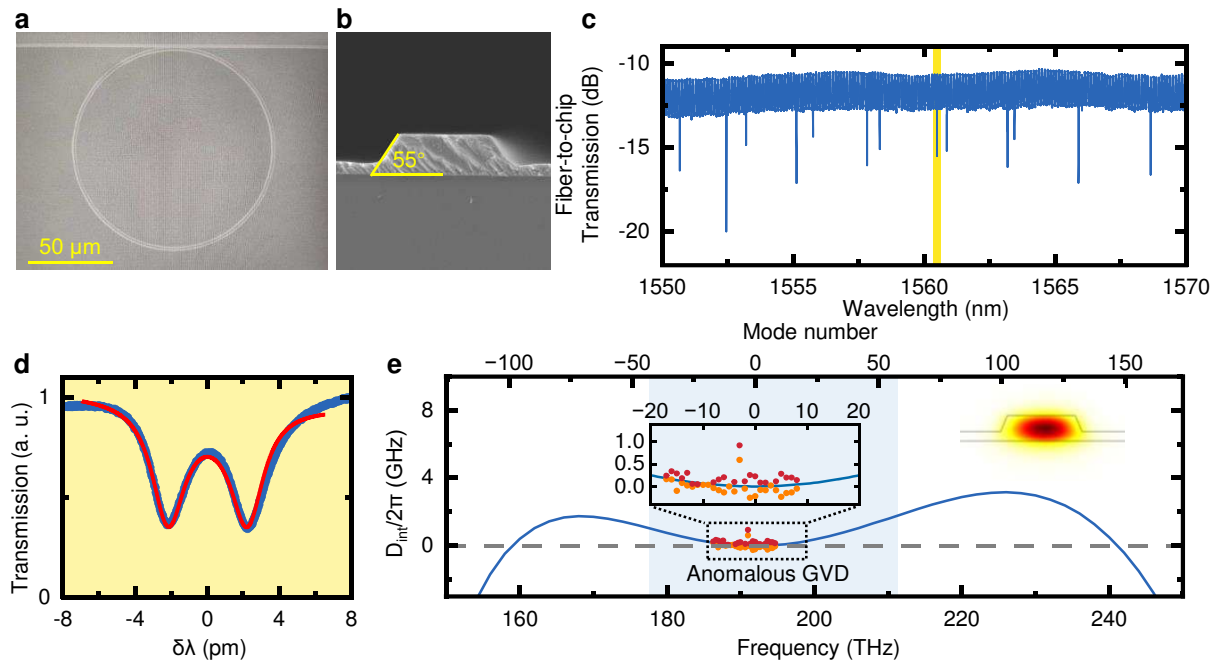}\caption{\label{fig:1}
\textbf{a}. Scanning electron microscopy (SEM) image of the device.
The air-cladded LiNbO$_3$ microring resonator has a radius of 60 $\mu$m.
\textbf{b}. SEM image of the cross section of the waveguide, showing a sidewall angle of $55{^\circ}$.
\textbf{c}. The typical transmission spectrum of the TE polarization in the microring.
\textbf{d}. A TE$_{00}$ resonance with mode split in \textbf{c}.
It has a linewidth of 260 MHz according to the fit (red curve), corresponding to a loaded $Q=7.38\times10^{5}$.
\textbf{e.} Integrated microresonator dispersion curve obtained by numerical simulation based on the actual device size.
Red (orange) dots represent the experimental results of the higher (lower) frequency of the splitting modes.}
\end{figure*}

In addition, a simple method to generate solitons is required for field-deployable applications.
Conventionally, access to the soliton state requires the pump laser to scan across the resonance from the blue-detuned side and settle on the red-detuned side \cite{Herr:14}.
However, due to the light absorption and thermal-optic effects of the microresonator, the abrupt power drop from the chaotic state to the soliton state leads to rapid cooling of the microresonator and resonance recoil, which immediately quench the soliton state \cite{Brasch:16, Li:17}.
Various sophisticated approaches to manage thermal effects have been developed, such as power kicking \cite{Brasch:16, Yi:16b}, single-sideband suppressed-carrier frequency shifters \cite{Stone:18}, dual-laser pump \cite{Zhou:19, LuZ:21} and pump modulation \cite{Wildi:19,Nishimoto:22}.
A recent breakthrough is the demonstration that, using laser self-injection locking \cite{Liang:15, Pavlov:18, Kondratiev:23},  solitons can be generated deterministically with a simple ``turnkey'' operation \cite{Shen:20}.
In this mode, soliton is generated on-demand by simple laser power-on, and maintains for hours without any active feedback control.
However, there is still one challenge with this method.
Simulation and experimental studies \cite{Voloshin:21, Briles:21} have shown that, nonlinear laser self-injection locking constrains the maximum attainable soliton detuning value. Therefore, single solitons, particularly the ones with octave spanning, are not feasible.

Here, we overcome the above challenges and demonstrate a self-emerging, broadband soliton with dual dispersive emission in an integrated lithium niobate (LiNbO$_3$) optical microresonator.
We harness the photorefractive effect of LiNbO$_3$ \cite{Savchenkov:06, Leidinger:16, Jiang:17} -- which is often considered a negative effect \cite{XuY:21} -- to significantly extend the soliton existence range.
Consequently, we successfully demonstrate a spontaneous yet deterministic single-soliton formation.


\begin{figure*}
\centering{}
\includegraphics[clip,width=16cm]{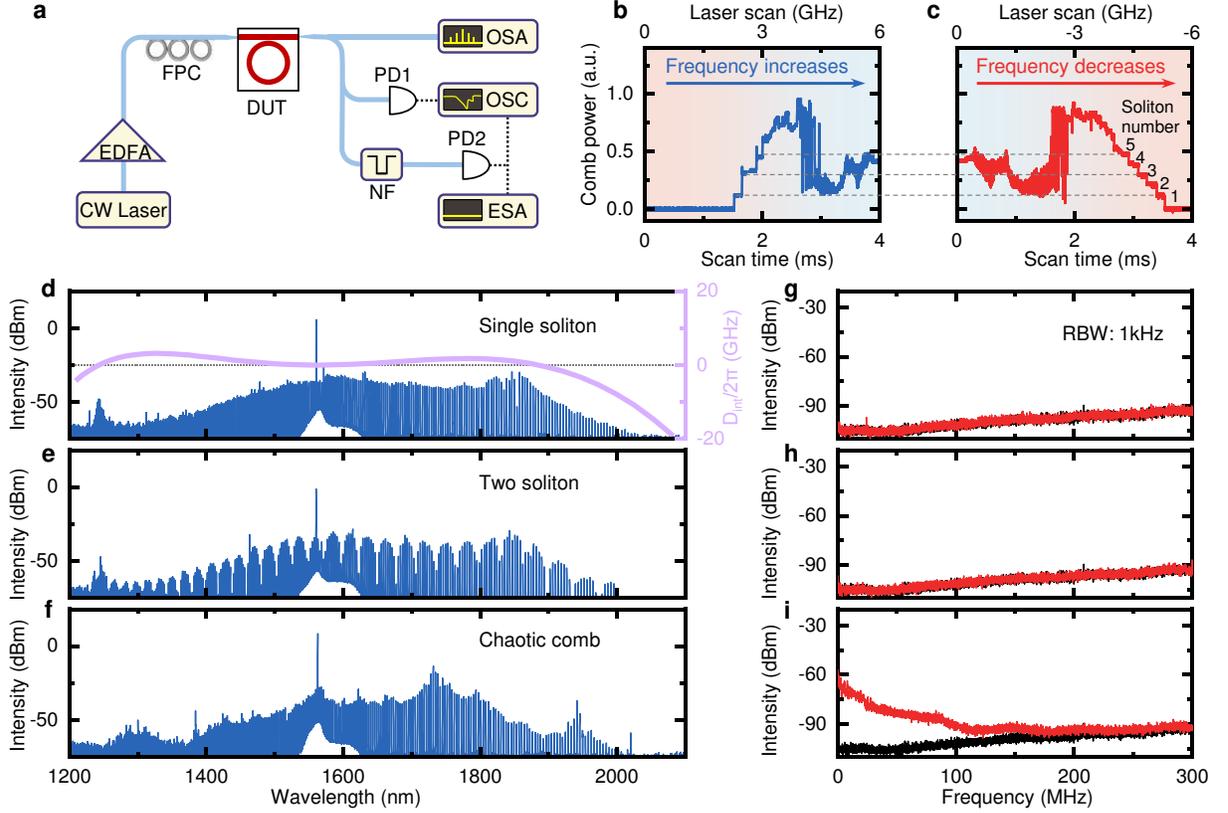}
\caption{
\label{fig:2}
\textbf{a}. Experimental setup for soliton generation.
EDFA, erbium-doped fiber amplifier;
FPC, fiber polarizationcontroller;
DUT, device under test;
NF, notch filter;
PD, photo detector;
OSA, optical spectrum analyzer;
OSC, oscilloscope;
ESA, electrical spectrum analyzer.
\textbf{b-c}. Comb powers evolution as the pump laser's frequency scans back and forth repeatedly in the red-detuned side of the resonance mode near 1560.6 nm. \textbf{d-f}. Optical spectra of the single-soliton,the two-soliton, and the chaotic comb states.
\textbf{g-i}: Low-frequency radio-frequency (RF) noise spectra (red) corresponding to \textbf{(d-f}).
Black lines are the noise spectra of detector.}
\end{figure*}

\section{Device design and soliton generation}

Our integrated microresonator is fabricated on a Z-cut, 580-nm-thick, LiNbO$_3$-on-insulator wafer (NANOLN).
Among a myriad of integrated material platforms developed for soliton microcombs \cite{Kovach:20}, LiNbO$_3$ \cite{Boes:18, Zhu:21} is particularly promising since it possesses both the Kerr and $\chi^2$ nonlinearities.
While the Kerr nonlinearity allows LiNbO$_3$ for soliton microcombs generation \cite{He:19, Bruch:21}, the $\chi^2$ nonlinearity makes LiNbO$_3$ ideal for electro-optic modulation \cite{HeM:19, WangC:18}. 
This unique combination makes LiNbO$_3$ ideal for fast tuning and modulation of soliton microcombs on a monolithic substrate \cite{Wang:19, Fang:19}.


The scanning electron microscopy (SEM) image of the device is shown in Fig. \ref{fig:1}a.
The air-cladded LN microring resonator has a radius of 60 $\mu$m, corresponding to an FSR of 334 GHz.
As illustrated in Fig. \ref{fig:1}b, the cross section of the microring resonator is trapezoidal, with a top width of 1.45 $\mu$m, an etching depth of 360 nm and a sidewall angle of $55{^\circ}$, leaving an LiNbO$_3$ slab of 220 nm.
Details of the fabrication process are provided in the Methods.
The transmission spectrum of TE modes is shown in Fig.\ref{fig:1}c, where the fundamental TE$_{00}$ mode and higher-order TE$_{10}$ mode can be observed.
A TE$_{00}$ resonance at 1560.6 nm is shown in Fig. \ref{fig:1}d.
The resonance exhibits a mode split profile, which is usually induced by the roughness of the resonators. 
A doublet model is used to fit the resonance (red curve) \cite{Gorodetsky:00}, showing a linewidth of 260 MHz, corresponding to a loaded $Q=7.38\times10^{5}$.


Anomalous microresonator dispersion is mandatory for soliton formation \cite{Herr:14}.
The integrated microresonator dispersion $D_\text{int}$ obtained by numerical simulation based on our actual device geometry is shown in Fig. \ref{fig:1}e.
Two phase-matching points, i.e. $D_\text{int}=0$ at approximately 160 THz and 240 THz on the simulated curve, indicate the emission of dispersive waves \cite{Milian:14, Jang:14, Brasch:15}, that can coherently extend the soliton spectral bandwidth \cite{Yu:19, LiuX:21, Gong:20}
Dots in Fig. \ref{fig:1}e show the measured $D_\text{int}$, where red (orange) dots represent the higher (lower) frequency dips of the split resonances.
The experimental results agrees with the simulated curve with an extracted $D_2/2\pi\sim1.4$ MHz.
The inset of Fig. \ref{fig:1}e illustrates the simulated optical modal profile.

A schematic of the experimental setup for soliton generation is shown in Fig. \ref{fig:2}a.
A pump power of 600 mW is coupled into the chip, while the on-chip pump power is around 150 mW.
Due to the photorefractive effect of LiNbO$_3$ \cite{Savchenkov:06, Leidinger:16, Jiang:17, XuY:21}, bi-directional switching of soliton states is feasible \cite{He:19}, as shown in Fig. \ref{fig:2}(b-c).
When the laser frequency scans back and forth repeatedly in the red-detuned side of the resonance mode near 1560.6 nm, the comb power shows a series of discrete steps, indicating a gradual change in the number of solitons.
The intracavity soliton number can be inferred from the step height.
Therefore, soliton states with different numbers can be deterministically accessed via simple laser frequency tuning to the corresponding soliton step.
Figure \ref{fig:2}d shows the optical spectrum of the single soliton state when the pump laser is stopped at the soliton step.
Two dispersive waves are observed at 1242 nm ($\sim241$ THz) and 1851 nm ($\sim162$ THz), consistent with the predicted $D_\text{int}$ curve (purple curve).
By increasing the laser frequency, a two-soliton state is observed, as shown in Fig. \ref{fig:2}e.
Further increase of the laser frequency to the resonance frequency, the soliton state collapses into a noisy, chaotic state.
Such a state is synergized by both the Kerr nonlinearity and Raman scattering \cite{YuM:20}, as shown in Fig. \ref{fig:2}f.
Figure \ref{fig:2}(g-i) show the low-frequency radio-frequency (RF) noise spectra with a resolution bandwidth of 1 kHz, which correspond to the comb states in Fig. \ref{fig:2}(d-f).
The flat spectra extending to DC in Fig. \ref{fig:2}(g, h) validate the coherent nature of soliton states.

\section{Soliton existence range}

\begin{figure*}
\begin{centering}
\includegraphics[clip,width=16cm]{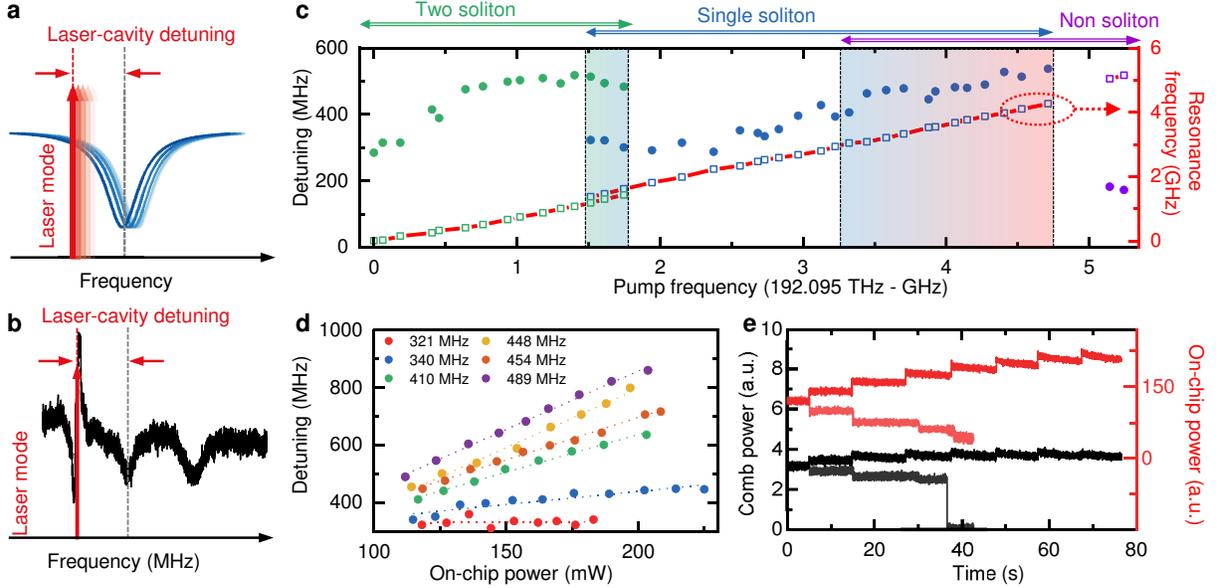}
\par\end{centering}
\centering{}\caption{\label{fig3}
\textbf{a}. Schematic diagram of the photorefraction-induced locking of the resonance frequency (blue) and laser frequency (red) during stable soliton operation.
\textbf{b}. The resonance at 1560.6 nm used in the experiment.
\textbf{c}. Variation in the resonance frequency (hollow squares) and laser-cavity detuning (solid dots) versus variation of the pump frequency.
\textbf{d}. Variation of the laser-cavity detuning with the changes of the on-chip pump power.
Different colors indicate different initial laser-cavity detunings.
\textbf{ e}. Variation of the transmitted power and comb power with the gradual change of the pump power.
Each step of transmitted power represents a pump power change of 20 mW.}
\end{figure*}

Next, we explore the influence of the photorefractive effect on the soliton formation
In addition to enable bi-directional access to soliton states, the photorefractive effect can also effectively extend the soliton existence range, as illustrated in Fig. \ref{fig3}a.
When the pump laser is red-detuned to the resonance, the resonance frequency is shifted and locked to the laser frequency due to the slow photorefraction of the LiNbO$_3$ microresonator.
This locking suppresses the variation of the effective laser-cavity detuning,  and thus extends the soliton existence range.
A weak probe light is employed to track the effective laser-cavity detuning in real time \cite{Niu:23,LiJ:22}.
As shown in Fig. \ref{fig3}b, the resonance at approximately 1560.6 nm is used in the experiment, and the position of the pump laser is indicated by the interference, which is marked by the red arrow.

Figure \ref{fig3}c shows the variation in the resonance frequency and laser-cavity detuning as the pump frequency gradually decreases.
Here we set the interval of the laser frequency tuning to 125 MHz ($\sim1$ pm), and the laser frequency is measured by a wavemeter.
Each value of laser-cavity detuning is obtained until the system is stable.
The initial laser frequency is 192.095 THz ($\sim1560.645$ nm), where the system is in a two-soliton state.
As the laser frequency decreases, the resonance frequency also decreases due to the photorefractive effect, as the red lines shown in Fig. \ref{fig3}c.
It is noted that the shift rate of the resonance frequency is slightly smaller than that of the laser frequency.
Therefore, with decreasing laser frequency, the laser-cavity detuning gradually increases.
When the laser frequency decreases by 1.5 GHz, the system begins to alternate between the two-soliton state and the single-soliton state, as the blue box shown in Fig. \ref{fig3}c.
When the laser frequency decreases by 1.94 GHz, the two-soliton state can no longer exist and the system stays in the single-soliton state.
In the range of laser frequency shift of 1.28 GHz, the single-soliton state is stable while the laser cavity-detuning is shifted by 135 MHz.  
Clearly, the intracavity state starts to alternate between the single-soliton and the non-soliton state when the laser frequency decreases by 3.22 GHz.
Until the pump frequency decreases by 5.14 GHz, the intracavity power can no longer allow soliton formation, and the system stays in a non-soliton state.
In this state, due to the further reduction of the intracavity power, the detuning is significantly reduced compared with that in the soliton state.

In addition, we further explore the stable existence range of the single-soliton with varied pump power.
Figure \ref{fig3}d shows the variation of the laser-cavity detuning of the single-soliton state with increasing on-chip pump power.
Different colors indicate different initial laser-cavity detuning values.
The on-chip pump power increases at intervals of 20 mW starting from 120 mW.
It clearly shows that the laser-cavity detuning increases with increasing on-chip pump power, and the greater value of the initial detuning provides greater rate of linear detuning changes with the on-chip pump power.
Figure \ref{fig3}e shows the comb power when the on-chip power is varied around 120 mW with an initial detuning of 400 MHz.
When the on-chip pump power drops below 60 mW, the comb power suddenly drops to zero, indicating that the single-soliton state vanishes.
When the pump power increases, the on-chip power will be less than expected due to the instability of the fiber-to-chip edge coupling.
According to the transmission power, when the on-chip power reaches 220 mW, the single-soliton state still exists.
As the pump power continues to increase, the on-chip power hardly changes due to the influence of edge coupling.
Therefore, it can be concluded that the single-soliton state is stable at least between 60 mW and 220 mW.
The detailed studies of the perturbation resistance of the solitons and their long-term stability are shown in Supplementary Information.
The results show that when the soliton state is perturbed or even disrupted, it can recover to the initial state  because of the photorefractive effect.
The soliton state can be maintained for 13 hours without any active control. 

\section{Turnkey generation of soliton microcomb}

\begin{figure}
\includegraphics[clip,width=1\columnwidth]{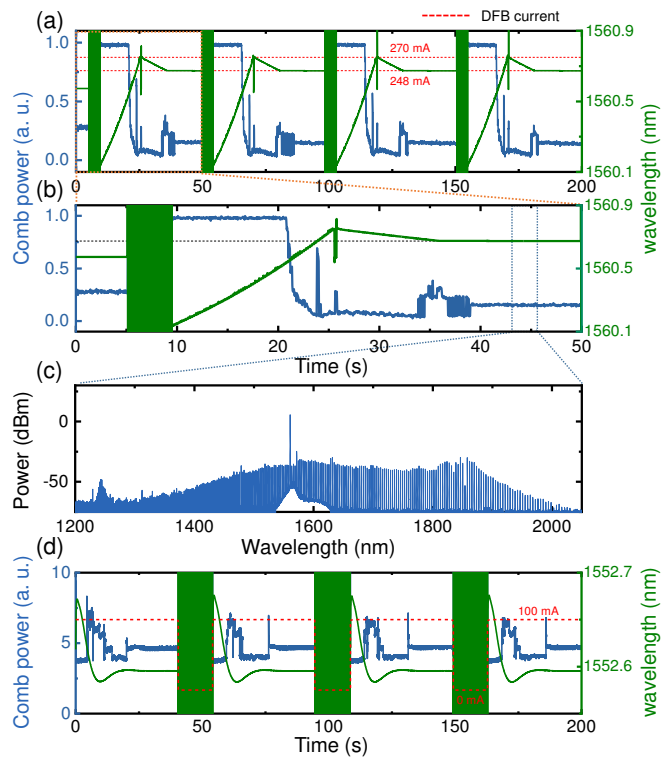}\caption{\label{fig4}
\textbf{a}. Reproducible turnkey soliton generation.
The DFB current rises from 0 mA to 270 mA and then drops to 248 mA.
The wavelength of 1560.675 nm corresponding to a current of 248 mA is within the single-soliton existence range shown in \textbf{c}.
\textbf{b}. Detailed generation and stabilization of the same single soliton without any active control.
\textbf{c}. Optical spectrum of the single soliton.
\textbf{d}. Repeat generation of the same single-soliton state.}
\end{figure}

As described above, the single-soliton state can be deterministically accessed without complex techniques.
Therefore, next, we use a DFB laser instead of a commercial tunable laser as the pump laser.
We control the laser frequency through the DFB current with a pre-set program.
The DFB parameters used in the experiment are listed in Fig.  \ref{fig4}.
Critical to our experiment are two current values (270 mA and 248 mA) that set the initial wavelength near the cold cavity mode (1560.753 nm) and the target wavelength (1560.675 nm) within the soliton existence range, respectively, as shown in Fig. \ref{fig4}a.
The DFB current rises to 270 mA in 17 seconds, and then drops to 248 mA in 10 seconds.
At the beginning, the current is low and the power is insufficient for the measurement by wavemeter, leading an oscillated and disordered signal.
As shown in Fig. \ref{fig4}b, this allows the pump laser to pass through the cold resonance mode and stabilize at the wavelength where the stable single-soliton state can be generated.
For the comb power, it finally stabilizes at the single-soliton power after a short period of oscillation.
As long as the primary current values are set, the system consistently accesses to the same single-soliton state regardless of the rise and fall time of the DFB current (see Supplementary Information).
By running the program repeatedly, as shown in Fig. \ref{fig4}a, the same single-soliton state can be repeatedly generated and stabilized without any active control.
The optical spectrum of the stable single-soliton state is shown in Fig. \ref{fig4}c.
The system can also operate in the two-soliton state or other soliton states with different DFB current values.

In addition, when the increasing speed of the DFB current is not controlled and the current rises instantaneously, due to the lag of the temperature change of the DFB laser, the actual wavelength of DFB will exceed the wavelength corresponding to the current value and then gradually fall back to the corresponding wavelength.
To exploit this behaviour, only one current value needs to be set at the wavelength where the soliton state is stable.
The DFB laser can enter the cold cavity mode by temperature flow.
Figure \ref{fig4}d shows repeated generation of the same single-soliton state using this method.
The stable pump wavelength of the soliton is 1552.59 nm.
Due to the increase in DFB temperature, the actual wavelength of DFB will rise to 1552.67 nm and then fall back.
When falling back, it passes through the resonance wavelength 1552.62 nm,  and blue-shifts the resonance to reach the stable soliton state.

\section{Conclusion}

In conclusion, we have demonstrated soliton self-emergence in a dispersion-optimized, integrated LiNbO$_3$ optical microresonator.
The single soliton features broad bandwidth with dual dispersive waves, ideal for $2f-3f$ self-referencing.
We have revealed that, via harnessing the photorefractive effect of LiNbO$_3$, the soliton existence range is significantly extended.
In contrast to the previous schemes where the laser frequency shift and the laser-cavity detuning shift are roughly the same \cite{Guo:16}, here the soliton detuning shift in LiNbO$_3$ is one-tenth of the laser frequency shift due to the photorefractive effect.
The single soliton is operated stably in a laser frequency range of 1.28 GHz and a on-chip power range from 60 mW to 220 mW, and is immune to frequency and power perturbations.
We show the system ability to recover from perturbation, long-term operation up to 13 hours, and turnkey soliton generation with a pre-programed DFB laser.
With further improvement of microresonator $Q$ and hybrid integration with DFB laser chips \cite{Shams-Ansari:22, LiM:22, Snigirev:23}, compact soliton microcomb devices with electronic actuation can be created.
These devices with robustness, stability, and portability can become central elements for LiDAR, microwave photonics and optical telecommunications.  

\vbox{}

\noindent\textbf{Methods}\\

\paragraph*{Device fabrication}

The device was fabricated on a 580-nm-thick, Z-cut LiNbO$_3$ thin film bonding to a silicon substrate with 500-$\mu$m-thick silicon and 2-$\mu$m-thick wet oxidation silicon dioxide (SiO$_2$).
Electron-beam lithography was used to pattern the device with hydrogen-silsesquioxane (HSQ) resist.
Following the development of the pattern, the film was partially etched by argon-ion-based reactive ion etching in an inductively coupled plasma (ICP) etcher to form a 360-nm-depth trapezoidal waveguide cross section with a remaining slab of 220 nm, as shown in Fig. \ref{fig:1}b.
The device was then cleaned by buffered HF solution and RCA1 cleaning solution (NH3:H2O2:H2O = 1:1:5) to remove the remaining HSQ resist and the redeposition formed in the etching process.
Finally, the facet of the chip was diced to enable good fiber-to-chip light coupling.

\paragraph*{Soliton generation}

A tunable laser (Toptica CTL) is employed as the pump laser, whose power is amplified by an erbium-doped fiber amplifier (EDFA).
The amplified light is then passed through the fiber polarization controller (FPC) to adjust the polarization state of the light to match the waveguide's TE mode  and is eventually launched onto the chip via a lensed fiber.
The fiber-to-chip coupling loss is approximately 6 dB per facet.
The output light is collected by another lensed fiber and split into three parts.
Forty-five percent of the light is sent into an optical spectrum analyzer (OSA) to characterize the microcomb spectrum.
Ten percent of the light is recorded by a photodetector (PD1) to monitor the transmission spectrum of the pump light.
The remaining light passes through a fiber Bragg grating (FBG) to attenuate the power of the pump component and is then recorded by another photodetector (PD2) to monitor the comb power and measure the low-frequency noise spectrum, respectively.

\paragraph*{Integrated dispersion}

The integrated dispersion, defined as $D_\text{int}=\omega_{\mu}-\omega_{0}-\mu D_{1}$, determines the spectrum width of the soliton microcomb.
Here, $\mu$ is the mode number, $\omega_0$ is the frequency of the reference resonance mode, $D_1$ is the FSR of the reference mode and $\omega_\mu$ is the frequency of the \textgreek{m}-th mode relative to the reference mode.
The red and orange dots in Fig. \ref{fig:1}f indicate the $D_\text{int}$ value measured by calibrating the transmission spectrum around the telecom band with the fiber Mach-Zehnder interferometer.

\vbox{}

\noindent\textbf{Data availability}\\
All data generated or analyzed during this study are available within the paper and its Supplementary Information.
Further source data will be made available on request.

\vbox{}

\noindent\textbf{Code availability}\\
The code used to solve the equations presented in the Supplementary Information will be made available on request.

\bibliographystyle{Microcavity}
\phantomsection\addcontentsline{toc}{section}{\refname}


\begin{thebibliography}{67}%
\makeatletter
\providecommand \@ifxundefined [1]{%
 \@ifx{#1\undefined}
}%
\providecommand \@ifnum [1]{%
 \ifnum #1\expandafter \@firstoftwo
 \else \expandafter \@secondoftwo
 \fi
}%
\providecommand \@ifx [1]{%
 \ifx #1\expandafter \@firstoftwo
 \else \expandafter \@secondoftwo
 \fi
}%
\providecommand \natexlab [1]{#1}%
\providecommand \enquote  [1]{``#1''}%
\providecommand \bibnamefont  [1]{#1}%
\providecommand \bibfnamefont [1]{#1}%
\providecommand \citenamefont [1]{#1}%
\providecommand \href@noop [0]{\@secondoftwo}%
\providecommand \href [0]{\begingroup \@sanitize@url \@href}%
\providecommand \@href[1]{\@@startlink{#1}\@@href}%
\providecommand \@@href[1]{\endgroup#1\@@endlink}%
\providecommand \@sanitize@url [0]{\catcode `\\12\catcode `\$12\catcode
  `\&12\catcode `\#12\catcode `\^12\catcode `\_12\catcode `\%12\relax}%
\providecommand \@@startlink[1]{}%
\providecommand \@@endlink[0]{}%
\providecommand \url  [0]{\begingroup\@sanitize@url \@url }%
\providecommand \@url [1]{\endgroup\@href {#1}{\urlprefix }}%
\providecommand \urlprefix  [0]{URL }%
\providecommand \Eprint [0]{\href }%
\providecommand \doibase [0]{http://dx.doi.org/}%
\providecommand \selectlanguage [0]{\@gobble}%
\providecommand \bibinfo  [0]{\@secondoftwo}%
\providecommand \bibfield  [0]{\@secondoftwo}%
\providecommand \translation [1]{[#1]}%
\providecommand \BibitemOpen [0]{}%
\providecommand \bibitemStop [0]{}%
\providecommand \bibitemNoStop [0]{.\EOS\space}%
\providecommand \EOS [0]{\spacefactor3000\relax}%
\providecommand \BibitemShut  [1]{\csname bibitem#1\endcsname}%
\let\auto@bib@innerbib\@empty
\bibitem [{\citenamefont {Leo}\ \emph {et~al.}(2010)\citenamefont {Leo},
  \citenamefont {Coen}, \citenamefont {Kockaert}, \citenamefont {Gorza},
  \citenamefont {Emplit},\ and\ \citenamefont {Haelterman}}]{Leo:10}%
  \BibitemOpen
  \bibfield  {author} {\bibinfo {author} {\bibfnamefont {F.}~\bibnamefont
  {Leo}}, \bibinfo {author} {\bibfnamefont {S.}~\bibnamefont {Coen}}, \bibinfo
  {author} {\bibfnamefont {P.}~\bibnamefont {Kockaert}}, \bibinfo {author}
  {\bibfnamefont {S.-P.}\ \bibnamefont {Gorza}}, \bibinfo {author}
  {\bibfnamefont {P.}~\bibnamefont {Emplit}}, \ and\ \bibinfo {author}
  {\bibfnamefont {M.}~\bibnamefont {Haelterman}},\ }\bibfield  {title}
  {\enquote {\bibinfo {title} {Temporal cavity solitons in one-dimensional kerr
  media as bits in an all-optical buffer},}\ }\href {\doibase
  10.1038/nphoton.2010.120} {\bibfield  {journal} {\bibinfo  {journal} {Nature
  Photonics}\ }\textbf {\bibinfo {volume} {4}},\ \bibinfo {pages} {471}
  (\bibinfo {year} {2010})}\BibitemShut {NoStop}%
\bibitem [{\citenamefont {Herr}\ \emph {et~al.}(2013)\citenamefont {Herr},
  \citenamefont {Brasch}, \citenamefont {Jost}, \citenamefont {Wang},
  \citenamefont {Kondratiev}, \citenamefont {Gorodetsky},\ and\ \citenamefont
  {Kippenberg}}]{Herr:14}%
  \BibitemOpen
  \bibfield  {author} {\bibinfo {author} {\bibfnamefont {T.}~\bibnamefont
  {Herr}}, \bibinfo {author} {\bibfnamefont {V.}~\bibnamefont {Brasch}},
  \bibinfo {author} {\bibfnamefont {J.~D.}\ \bibnamefont {Jost}}, \bibinfo
  {author} {\bibfnamefont {C.~Y.}\ \bibnamefont {Wang}}, \bibinfo {author}
  {\bibfnamefont {N.~M.}\ \bibnamefont {Kondratiev}}, \bibinfo {author}
  {\bibfnamefont {M.~L.}\ \bibnamefont {Gorodetsky}}, \ and\ \bibinfo {author}
  {\bibfnamefont {T.~J.}\ \bibnamefont {Kippenberg}},\ }\bibfield  {title}
  {\enquote {\bibinfo {title} {Temporal solitons in optical microresonators},}\
  }\href {https://doi.org/10.1038/nphoton.2013.343} {\bibfield  {journal}
  {\bibinfo  {journal} {Nature Photonics}\ }\textbf {\bibinfo {volume} {8}},\
  \bibinfo {pages} {145} (\bibinfo {year} {2013})}\BibitemShut {NoStop}%
\bibitem [{\citenamefont {Kippenberg}\ \emph {et~al.}(2018)\citenamefont
  {Kippenberg}, \citenamefont {Gaeta}, \citenamefont {Lipson},\ and\
  \citenamefont {Gorodetsky}}]{Kippenberg:18}%
  \BibitemOpen
  \bibfield  {author} {\bibinfo {author} {\bibfnamefont {T.~J.}\ \bibnamefont
  {Kippenberg}}, \bibinfo {author} {\bibfnamefont {A.~L.}\ \bibnamefont
  {Gaeta}}, \bibinfo {author} {\bibfnamefont {M.}~\bibnamefont {Lipson}}, \
  and\ \bibinfo {author} {\bibfnamefont {M.~L.}\ \bibnamefont {Gorodetsky}},\
  }\bibfield  {title} {\enquote {\bibinfo {title} {Dissipative kerr solitons in
  optical microresonators},}\ }\href {\doibase 10.1126/science.aan8083}
  {\bibfield  {journal} {\bibinfo  {journal} {Science}\ }\textbf {\bibinfo
  {volume} {361}},\ \bibinfo {pages} {eaan8083} (\bibinfo {year}
  {2018})}\BibitemShut {NoStop}%
\bibitem [{\citenamefont {Pasquazi}\ \emph {et~al.}(2018)\citenamefont
  {Pasquazi}, \citenamefont {Peccianti}, \citenamefont {Razzari}, \citenamefont
  {Moss}, \citenamefont {Coen}, \citenamefont {Erkintalo}, \citenamefont
  {Chembo}, \citenamefont {Hansson}, \citenamefont {Wabnitz}, \citenamefont
  {Del'Haye}, \citenamefont {Xue}, \citenamefont {Weiner},\ and\ \citenamefont
  {Morandotti}}]{Pasquazi:18}%
  \BibitemOpen
  \bibfield  {author} {\bibinfo {author} {\bibfnamefont {A.}~\bibnamefont
  {Pasquazi}}, \bibinfo {author} {\bibfnamefont {M.}~\bibnamefont {Peccianti}},
  \bibinfo {author} {\bibfnamefont {L.}~\bibnamefont {Razzari}}, \bibinfo
  {author} {\bibfnamefont {D.~J.}\ \bibnamefont {Moss}}, \bibinfo {author}
  {\bibfnamefont {S.}~\bibnamefont {Coen}}, \bibinfo {author} {\bibfnamefont
  {M.}~\bibnamefont {Erkintalo}}, \bibinfo {author} {\bibfnamefont {Y.~K.}\
  \bibnamefont {Chembo}}, \bibinfo {author} {\bibfnamefont {T.}~\bibnamefont
  {Hansson}}, \bibinfo {author} {\bibfnamefont {S.}~\bibnamefont {Wabnitz}},
  \bibinfo {author} {\bibfnamefont {P.}~\bibnamefont {Del'Haye}}, \bibinfo
  {author} {\bibfnamefont {X.}~\bibnamefont {Xue}}, \bibinfo {author}
  {\bibfnamefont {A.~M.}\ \bibnamefont {Weiner}}, \ and\ \bibinfo {author}
  {\bibfnamefont {R.}~\bibnamefont {Morandotti}},\ }\bibfield  {title}
  {\enquote {\bibinfo {title} {Micro-combs: A novel generation of optical
  sources},}\ }\bibfield  {booktitle} {\emph {\bibinfo {booktitle}
  {Micro-combs: A novel generation of optical sources}},\ }\href {\doibase
  https://doi.org/10.1016/j.physrep.2017.08.004} {\bibfield  {journal}
  {\bibinfo  {journal} {Physics Reports}\ }\textbf {\bibinfo {volume} {729}},\
  \bibinfo {pages} {1} (\bibinfo {year} {2018})}\BibitemShut {NoStop}%
\bibitem [{\citenamefont {Yi}\ \emph {et~al.}(2015)\citenamefont {Yi},
  \citenamefont {Yang}, \citenamefont {Yang}, \citenamefont {Suh},\ and\
  \citenamefont {Vahala}}]{Yi:15}%
  \BibitemOpen
  \bibfield  {author} {\bibinfo {author} {\bibfnamefont {X.}~\bibnamefont
  {Yi}}, \bibinfo {author} {\bibfnamefont {Q.-F.}\ \bibnamefont {Yang}},
  \bibinfo {author} {\bibfnamefont {K.~Y.}\ \bibnamefont {Yang}}, \bibinfo
  {author} {\bibfnamefont {M.-G.}\ \bibnamefont {Suh}}, \ and\ \bibinfo
  {author} {\bibfnamefont {K.}~\bibnamefont {Vahala}},\ }\bibfield  {title}
  {\enquote {\bibinfo {title} {Soliton frequency comb at microwave rates in a
  high-q silica microresonator},}\ }\href {\doibase 10.1364/OPTICA.2.001078}
  {\bibfield  {journal} {\bibinfo  {journal} {Optica}\ }\textbf {\bibinfo
  {volume} {2}},\ \bibinfo {pages} {1078} (\bibinfo {year} {2015})}\BibitemShut
  {NoStop}%
\bibitem [{\citenamefont {Suh}\ and\ \citenamefont
  {Vahala}(2018{\natexlab{a}})}]{Suh:18b}%
  \BibitemOpen
  \bibfield  {author} {\bibinfo {author} {\bibfnamefont {M.-G.}\ \bibnamefont
  {Suh}}\ and\ \bibinfo {author} {\bibfnamefont {K.}~\bibnamefont {Vahala}},\
  }\bibfield  {title} {\enquote {\bibinfo {title} {Gigahertz-repetition-rate
  soliton microcombs},}\ }\href {\doibase 10.1364/OPTICA.5.000065} {\bibfield
  {journal} {\bibinfo  {journal} {Optica}\ }\textbf {\bibinfo {volume} {5}},\
  \bibinfo {pages} {65} (\bibinfo {year} {2018}{\natexlab{a}})}\BibitemShut
  {NoStop}%
\bibitem [{\citenamefont {Pfeiffer}\ \emph {et~al.}(2017)\citenamefont
  {Pfeiffer}, \citenamefont {Herkommer}, \citenamefont {Liu}, \citenamefont
  {Guo}, \citenamefont {Karpov}, \citenamefont {Lucas}, \citenamefont
  {Zervas},\ and\ \citenamefont {Kippenberg}}]{Pfeiffer:17}%
  \BibitemOpen
  \bibfield  {author} {\bibinfo {author} {\bibfnamefont {M.~H.~P.}\
  \bibnamefont {Pfeiffer}}, \bibinfo {author} {\bibfnamefont {C.}~\bibnamefont
  {Herkommer}}, \bibinfo {author} {\bibfnamefont {J.}~\bibnamefont {Liu}},
  \bibinfo {author} {\bibfnamefont {H.}~\bibnamefont {Guo}}, \bibinfo {author}
  {\bibfnamefont {M.}~\bibnamefont {Karpov}}, \bibinfo {author} {\bibfnamefont
  {E.}~\bibnamefont {Lucas}}, \bibinfo {author} {\bibfnamefont
  {M.}~\bibnamefont {Zervas}}, \ and\ \bibinfo {author} {\bibfnamefont {T.~J.}\
  \bibnamefont {Kippenberg}},\ }\bibfield  {title} {\enquote {\bibinfo {title}
  {Octave-spanning dissipative kerr soliton frequency combs in si3n4
  microresonators},}\ }\href {\doibase 10.1364/OPTICA.4.000684} {\bibfield
  {journal} {\bibinfo  {journal} {Optica}\ }\textbf {\bibinfo {volume} {4}},\
  \bibinfo {pages} {684} (\bibinfo {year} {2017})}\BibitemShut {NoStop}%
\bibitem [{\citenamefont {Li}\ \emph {et~al.}(2017)\citenamefont {Li},
  \citenamefont {Briles}, \citenamefont {Westly}, \citenamefont {Drake},
  \citenamefont {Stone}, \citenamefont {Ilic}, \citenamefont {Diddams},
  \citenamefont {Papp},\ and\ \citenamefont {Srinivasan}}]{Li:17}%
  \BibitemOpen
  \bibfield  {author} {\bibinfo {author} {\bibfnamefont {Q.}~\bibnamefont
  {Li}}, \bibinfo {author} {\bibfnamefont {T.~C.}\ \bibnamefont {Briles}},
  \bibinfo {author} {\bibfnamefont {D.~A.}\ \bibnamefont {Westly}}, \bibinfo
  {author} {\bibfnamefont {T.~E.}\ \bibnamefont {Drake}}, \bibinfo {author}
  {\bibfnamefont {J.~R.}\ \bibnamefont {Stone}}, \bibinfo {author}
  {\bibfnamefont {B.~R.}\ \bibnamefont {Ilic}}, \bibinfo {author}
  {\bibfnamefont {S.~A.}\ \bibnamefont {Diddams}}, \bibinfo {author}
  {\bibfnamefont {S.~B.}\ \bibnamefont {Papp}}, \ and\ \bibinfo {author}
  {\bibfnamefont {K.}~\bibnamefont {Srinivasan}},\ }\bibfield  {title}
  {\enquote {\bibinfo {title} {Stably accessing octave-spanning microresonator
  frequency combs in the soliton regime},}\ }\href {\doibase
  10.1364/OPTICA.4.000193} {\bibfield  {journal} {\bibinfo  {journal} {Optica}\
  }\textbf {\bibinfo {volume} {4}},\ \bibinfo {pages} {193} (\bibinfo {year}
  {2017})}\BibitemShut {NoStop}%
\bibitem [{\citenamefont {Moss}\ \emph {et~al.}(2013)\citenamefont {Moss},
  \citenamefont {Morandotti}, \citenamefont {Gaeta},\ and\ \citenamefont
  {Lipson}}]{Moss:13}%
  \BibitemOpen
  \bibfield  {author} {\bibinfo {author} {\bibfnamefont {D.~J.}\ \bibnamefont
  {Moss}}, \bibinfo {author} {\bibfnamefont {R.}~\bibnamefont {Morandotti}},
  \bibinfo {author} {\bibfnamefont {A.~L.}\ \bibnamefont {Gaeta}}, \ and\
  \bibinfo {author} {\bibfnamefont {M.}~\bibnamefont {Lipson}},\ }\bibfield
  {title} {\enquote {\bibinfo {title} {New cmos-compatible platforms based on
  silicon nitride and hydex for nonlinear optics},}\ }\href
  {https://doi.org/10.1038/nphoton.2013.183} {\bibfield  {journal} {\bibinfo
  {journal} {Nature Photonics}\ }\textbf {\bibinfo {volume} {7}},\ \bibinfo
  {pages} {597} (\bibinfo {year} {2013})}\BibitemShut {NoStop}%
\bibitem [{\citenamefont {Gaeta}\ \emph {et~al.}(2019)\citenamefont {Gaeta},
  \citenamefont {Lipson},\ and\ \citenamefont {Kippenberg}}]{Gaeta:19}%
  \BibitemOpen
  \bibfield  {author} {\bibinfo {author} {\bibfnamefont {A.~L.}\ \bibnamefont
  {Gaeta}}, \bibinfo {author} {\bibfnamefont {M.}~\bibnamefont {Lipson}}, \
  and\ \bibinfo {author} {\bibfnamefont {T.~J.}\ \bibnamefont {Kippenberg}},\
  }\bibfield  {title} {\enquote {\bibinfo {title} {Photonic-chip-based
  frequency combs},}\ }\href {\doibase 10.1038/s41566-019-0358-x} {\bibfield
  {journal} {\bibinfo  {journal} {Nature Photonics}\ }\textbf {\bibinfo
  {volume} {13}},\ \bibinfo {pages} {158} (\bibinfo {year} {2019})}\BibitemShut
  {NoStop}%
\bibitem [{\citenamefont {Kovach}\ \emph {et~al.}(2020)\citenamefont {Kovach},
  \citenamefont {Chen}, \citenamefont {He}, \citenamefont {Choi}, \citenamefont
  {Dogan}, \citenamefont {Ghasemkhani}, \citenamefont {Taheri},\ and\
  \citenamefont {Armani}}]{Kovach:20}%
  \BibitemOpen
  \bibfield  {author} {\bibinfo {author} {\bibfnamefont {A.}~\bibnamefont
  {Kovach}}, \bibinfo {author} {\bibfnamefont {D.}~\bibnamefont {Chen}},
  \bibinfo {author} {\bibfnamefont {J.}~\bibnamefont {He}}, \bibinfo {author}
  {\bibfnamefont {H.}~\bibnamefont {Choi}}, \bibinfo {author} {\bibfnamefont
  {A.~H.}\ \bibnamefont {Dogan}}, \bibinfo {author} {\bibfnamefont
  {M.}~\bibnamefont {Ghasemkhani}}, \bibinfo {author} {\bibfnamefont
  {H.}~\bibnamefont {Taheri}}, \ and\ \bibinfo {author} {\bibfnamefont {A.~M.}\
  \bibnamefont {Armani}},\ }\bibfield  {title} {\enquote {\bibinfo {title}
  {Emerging material systems for integrated optical kerr frequency combs},}\
  }\href {\doibase 10.1364/AOP.376924} {\bibfield  {journal} {\bibinfo
  {journal} {Adv. Opt. Photon.}\ }\textbf {\bibinfo {volume} {12}},\ \bibinfo
  {pages} {135} (\bibinfo {year} {2020})}\BibitemShut {NoStop}%
\bibitem [{\citenamefont {Marin-Palomo}\ \emph {et~al.}(2017)\citenamefont
  {Marin-Palomo}, \citenamefont {Kemal}, \citenamefont {Karpov}, \citenamefont
  {Kordts}, \citenamefont {Pfeifle}, \citenamefont {Pfeiffer}, \citenamefont
  {Trocha}, \citenamefont {Wolf}, \citenamefont {Brasch}, \citenamefont
  {Anderson}, \citenamefont {Rosenberger}, \citenamefont {Vijayan},
  \citenamefont {Freude}, \citenamefont {Kippenberg},\ and\ \citenamefont
  {Koos}}]{Marin-Palomo:17}%
  \BibitemOpen
  \bibfield  {author} {\bibinfo {author} {\bibfnamefont {P.}~\bibnamefont
  {Marin-Palomo}}, \bibinfo {author} {\bibfnamefont {J.~N.}\ \bibnamefont
  {Kemal}}, \bibinfo {author} {\bibfnamefont {M.}~\bibnamefont {Karpov}},
  \bibinfo {author} {\bibfnamefont {A.}~\bibnamefont {Kordts}}, \bibinfo
  {author} {\bibfnamefont {J.}~\bibnamefont {Pfeifle}}, \bibinfo {author}
  {\bibfnamefont {M.~H.~P.}\ \bibnamefont {Pfeiffer}}, \bibinfo {author}
  {\bibfnamefont {P.}~\bibnamefont {Trocha}}, \bibinfo {author} {\bibfnamefont
  {S.}~\bibnamefont {Wolf}}, \bibinfo {author} {\bibfnamefont {V.}~\bibnamefont
  {Brasch}}, \bibinfo {author} {\bibfnamefont {M.~H.}\ \bibnamefont
  {Anderson}}, \bibinfo {author} {\bibfnamefont {R.}~\bibnamefont
  {Rosenberger}}, \bibinfo {author} {\bibfnamefont {K.}~\bibnamefont
  {Vijayan}}, \bibinfo {author} {\bibfnamefont {W.}~\bibnamefont {Freude}},
  \bibinfo {author} {\bibfnamefont {T.~J.}\ \bibnamefont {Kippenberg}}, \ and\
  \bibinfo {author} {\bibfnamefont {C.}~\bibnamefont {Koos}},\ }\bibfield
  {title} {\enquote {\bibinfo {title} {Microresonator-based solitons for
  massively parallel coherent optical communications},}\ }\href
  {https://doi.org/10.1038/nature22387} {\bibfield  {journal} {\bibinfo
  {journal} {Nature}\ }\textbf {\bibinfo {volume} {546}},\ \bibinfo {pages}
  {274} (\bibinfo {year} {2017})}\BibitemShut {NoStop}%
\bibitem [{\citenamefont {Fujii}\ \emph {et~al.}(2022)\citenamefont {Fujii},
  \citenamefont {Tanaka}, \citenamefont {Ohtsuka}, \citenamefont {Kogure},
  \citenamefont {Wada}, \citenamefont {Kumazaki}, \citenamefont {Tasaka},
  \citenamefont {Hashimoto}, \citenamefont {Kobayashi}, \citenamefont {Araki},
  \citenamefont {Furusawa}, \citenamefont {Sekine}, \citenamefont {Kawanishi},\
  and\ \citenamefont {Tanabe}}]{Fujii:22}%
  \BibitemOpen
  \bibfield  {author} {\bibinfo {author} {\bibfnamefont {S.}~\bibnamefont
  {Fujii}}, \bibinfo {author} {\bibfnamefont {S.}~\bibnamefont {Tanaka}},
  \bibinfo {author} {\bibfnamefont {T.}~\bibnamefont {Ohtsuka}}, \bibinfo
  {author} {\bibfnamefont {S.}~\bibnamefont {Kogure}}, \bibinfo {author}
  {\bibfnamefont {K.}~\bibnamefont {Wada}}, \bibinfo {author} {\bibfnamefont
  {H.}~\bibnamefont {Kumazaki}}, \bibinfo {author} {\bibfnamefont
  {S.}~\bibnamefont {Tasaka}}, \bibinfo {author} {\bibfnamefont
  {Y.}~\bibnamefont {Hashimoto}}, \bibinfo {author} {\bibfnamefont
  {Y.}~\bibnamefont {Kobayashi}}, \bibinfo {author} {\bibfnamefont
  {T.}~\bibnamefont {Araki}}, \bibinfo {author} {\bibfnamefont
  {K.}~\bibnamefont {Furusawa}}, \bibinfo {author} {\bibfnamefont
  {N.}~\bibnamefont {Sekine}}, \bibinfo {author} {\bibfnamefont
  {S.}~\bibnamefont {Kawanishi}}, \ and\ \bibinfo {author} {\bibfnamefont
  {T.}~\bibnamefont {Tanabe}},\ }\bibfield  {title} {\enquote {\bibinfo {title}
  {Dissipative kerr soliton microcombs for fec-free optical communications over
  100 channels},}\ }\href {\doibase 10.1364/OE.447712} {\bibfield  {journal}
  {\bibinfo  {journal} {Opt. Express}\ }\textbf {\bibinfo {volume} {30}},\
  \bibinfo {pages} {1351} (\bibinfo {year} {2022})}\BibitemShut {NoStop}%
\bibitem [{\citenamefont {Obrzud}\ \emph {et~al.}(2019)\citenamefont {Obrzud},
  \citenamefont {Rainer}, \citenamefont {Harutyunyan}, \citenamefont
  {Anderson}, \citenamefont {Liu}, \citenamefont {Geiselmann}, \citenamefont
  {Chazelas}, \citenamefont {Kundermann}, \citenamefont {Lecomte},
  \citenamefont {Cecconi}, \citenamefont {Ghedina}, \citenamefont {Molinari},
  \citenamefont {Pepe}, \citenamefont {Wildi}, \citenamefont {Bouchy},
  \citenamefont {Kippenberg},\ and\ \citenamefont {Herr}}]{Obrzud:19}%
  \BibitemOpen
  \bibfield  {author} {\bibinfo {author} {\bibfnamefont {E.}~\bibnamefont
  {Obrzud}}, \bibinfo {author} {\bibfnamefont {M.}~\bibnamefont {Rainer}},
  \bibinfo {author} {\bibfnamefont {A.}~\bibnamefont {Harutyunyan}}, \bibinfo
  {author} {\bibfnamefont {M.~H.}\ \bibnamefont {Anderson}}, \bibinfo {author}
  {\bibfnamefont {J.}~\bibnamefont {Liu}}, \bibinfo {author} {\bibfnamefont
  {M.}~\bibnamefont {Geiselmann}}, \bibinfo {author} {\bibfnamefont
  {B.}~\bibnamefont {Chazelas}}, \bibinfo {author} {\bibfnamefont
  {S.}~\bibnamefont {Kundermann}}, \bibinfo {author} {\bibfnamefont
  {S.}~\bibnamefont {Lecomte}}, \bibinfo {author} {\bibfnamefont
  {M.}~\bibnamefont {Cecconi}}, \bibinfo {author} {\bibfnamefont
  {A.}~\bibnamefont {Ghedina}}, \bibinfo {author} {\bibfnamefont
  {E.}~\bibnamefont {Molinari}}, \bibinfo {author} {\bibfnamefont
  {F.}~\bibnamefont {Pepe}}, \bibinfo {author} {\bibfnamefont {F.}~\bibnamefont
  {Wildi}}, \bibinfo {author} {\bibfnamefont {F.}~\bibnamefont {Bouchy}},
  \bibinfo {author} {\bibfnamefont {T.~J.}\ \bibnamefont {Kippenberg}}, \ and\
  \bibinfo {author} {\bibfnamefont {T.}~\bibnamefont {Herr}},\ }\bibfield
  {title} {\enquote {\bibinfo {title} {A microphotonic astrocomb},}\ }\href
  {\doibase 10.1038/s41566-018-0309-y} {\bibfield  {journal} {\bibinfo
  {journal} {Nature Photonics}\ }\textbf {\bibinfo {volume} {13}},\ \bibinfo
  {pages} {31} (\bibinfo {year} {2019})}\BibitemShut {NoStop}%
\bibitem [{\citenamefont {Suh}\ \emph {et~al.}(2019)\citenamefont {Suh},
  \citenamefont {Yi}, \citenamefont {Lai}, \citenamefont {Leifer},
  \citenamefont {Grudinin}, \citenamefont {Vasisht}, \citenamefont {Martin},
  \citenamefont {Fitzgerald}, \citenamefont {Doppmann}, \citenamefont {Wang},
  \citenamefont {Mawet}, \citenamefont {Papp}, \citenamefont {Diddams},
  \citenamefont {Beichman},\ and\ \citenamefont {Vahala}}]{Suh:19}%
  \BibitemOpen
  \bibfield  {author} {\bibinfo {author} {\bibfnamefont {M.-G.}\ \bibnamefont
  {Suh}}, \bibinfo {author} {\bibfnamefont {X.}~\bibnamefont {Yi}}, \bibinfo
  {author} {\bibfnamefont {Y.-H.}\ \bibnamefont {Lai}}, \bibinfo {author}
  {\bibfnamefont {S.}~\bibnamefont {Leifer}}, \bibinfo {author} {\bibfnamefont
  {I.~S.}\ \bibnamefont {Grudinin}}, \bibinfo {author} {\bibfnamefont
  {G.}~\bibnamefont {Vasisht}}, \bibinfo {author} {\bibfnamefont {E.~C.}\
  \bibnamefont {Martin}}, \bibinfo {author} {\bibfnamefont {M.~P.}\
  \bibnamefont {Fitzgerald}}, \bibinfo {author} {\bibfnamefont
  {G.}~\bibnamefont {Doppmann}}, \bibinfo {author} {\bibfnamefont
  {J.}~\bibnamefont {Wang}}, \bibinfo {author} {\bibfnamefont {D.}~\bibnamefont
  {Mawet}}, \bibinfo {author} {\bibfnamefont {S.~B.}\ \bibnamefont {Papp}},
  \bibinfo {author} {\bibfnamefont {S.~A.}\ \bibnamefont {Diddams}}, \bibinfo
  {author} {\bibfnamefont {C.}~\bibnamefont {Beichman}}, \ and\ \bibinfo
  {author} {\bibfnamefont {K.}~\bibnamefont {Vahala}},\ }\bibfield  {title}
  {\enquote {\bibinfo {title} {Searching for exoplanets using a microresonator
  astrocomb},}\ }\href {\doibase 10.1038/s41566-018-0312-3} {\bibfield
  {journal} {\bibinfo  {journal} {Nature Photonics}\ }\textbf {\bibinfo
  {volume} {13}},\ \bibinfo {pages} {25} (\bibinfo {year} {2019})}\BibitemShut
  {NoStop}%
\bibitem [{\citenamefont {Trocha}\ \emph {et~al.}(2018)\citenamefont {Trocha},
  \citenamefont {Karpov}, \citenamefont {Ganin}, \citenamefont {Pfeiffer},
  \citenamefont {Kordts}, \citenamefont {Wolf}, \citenamefont {Krockenberger},
  \citenamefont {Marin-Palomo}, \citenamefont {Weimann}, \citenamefont
  {Randel}, \citenamefont {Freude}, \citenamefont {Kippenberg},\ and\
  \citenamefont {Koos}}]{Trocha:18}%
  \BibitemOpen
  \bibfield  {author} {\bibinfo {author} {\bibfnamefont {P.}~\bibnamefont
  {Trocha}}, \bibinfo {author} {\bibfnamefont {M.}~\bibnamefont {Karpov}},
  \bibinfo {author} {\bibfnamefont {D.}~\bibnamefont {Ganin}}, \bibinfo
  {author} {\bibfnamefont {M.~H.~P.}\ \bibnamefont {Pfeiffer}}, \bibinfo
  {author} {\bibfnamefont {A.}~\bibnamefont {Kordts}}, \bibinfo {author}
  {\bibfnamefont {S.}~\bibnamefont {Wolf}}, \bibinfo {author} {\bibfnamefont
  {J.}~\bibnamefont {Krockenberger}}, \bibinfo {author} {\bibfnamefont
  {P.}~\bibnamefont {Marin-Palomo}}, \bibinfo {author} {\bibfnamefont
  {C.}~\bibnamefont {Weimann}}, \bibinfo {author} {\bibfnamefont
  {S.}~\bibnamefont {Randel}}, \bibinfo {author} {\bibfnamefont
  {W.}~\bibnamefont {Freude}}, \bibinfo {author} {\bibfnamefont {T.~J.}\
  \bibnamefont {Kippenberg}}, \ and\ \bibinfo {author} {\bibfnamefont
  {C.}~\bibnamefont {Koos}},\ }\bibfield  {title} {\enquote {\bibinfo {title}
  {Ultrafast optical ranging using microresonator soliton frequency combs},}\
  }\href {\doibase 10.1126/science.aao3924} {\bibfield  {journal} {\bibinfo
  {journal} {Science}\ }\textbf {\bibinfo {volume} {359}},\ \bibinfo {pages}
  {887} (\bibinfo {year} {2018})}\BibitemShut {NoStop}%
\bibitem [{\citenamefont {Suh}\ and\ \citenamefont
  {Vahala}(2018{\natexlab{b}})}]{Suh:18}%
  \BibitemOpen
  \bibfield  {author} {\bibinfo {author} {\bibfnamefont {M.-G.}\ \bibnamefont
  {Suh}}\ and\ \bibinfo {author} {\bibfnamefont {K.~J.}\ \bibnamefont
  {Vahala}},\ }\bibfield  {title} {\enquote {\bibinfo {title} {Soliton
  microcomb range measurement},}\ }\href {\doibase 10.1126/science.aao1968}
  {\bibfield  {journal} {\bibinfo  {journal} {Science}\ }\textbf {\bibinfo
  {volume} {359}},\ \bibinfo {pages} {884} (\bibinfo {year}
  {2018}{\natexlab{b}})}\BibitemShut {NoStop}%
\bibitem [{\citenamefont {Feldmann}\ \emph {et~al.}(2021)\citenamefont
  {Feldmann}, \citenamefont {Youngblood}, \citenamefont {Karpov}, \citenamefont
  {Gehring}, \citenamefont {Li}, \citenamefont {Stappers}, \citenamefont
  {Le~Gallo}, \citenamefont {Fu}, \citenamefont {Lukashchuk}, \citenamefont
  {Raja}, \citenamefont {Liu}, \citenamefont {Wright}, \citenamefont
  {Sebastian}, \citenamefont {Kippenberg}, \citenamefont {Pernice},\ and\
  \citenamefont {Bhaskaran}}]{Feldmann:21}%
  \BibitemOpen
  \bibfield  {author} {\bibinfo {author} {\bibfnamefont {J.}~\bibnamefont
  {Feldmann}}, \bibinfo {author} {\bibfnamefont {N.}~\bibnamefont
  {Youngblood}}, \bibinfo {author} {\bibfnamefont {M.}~\bibnamefont {Karpov}},
  \bibinfo {author} {\bibfnamefont {H.}~\bibnamefont {Gehring}}, \bibinfo
  {author} {\bibfnamefont {X.}~\bibnamefont {Li}}, \bibinfo {author}
  {\bibfnamefont {M.}~\bibnamefont {Stappers}}, \bibinfo {author}
  {\bibfnamefont {M.}~\bibnamefont {Le~Gallo}}, \bibinfo {author}
  {\bibfnamefont {X.}~\bibnamefont {Fu}}, \bibinfo {author} {\bibfnamefont
  {A.}~\bibnamefont {Lukashchuk}}, \bibinfo {author} {\bibfnamefont {A.~S.}\
  \bibnamefont {Raja}}, \bibinfo {author} {\bibfnamefont {J.}~\bibnamefont
  {Liu}}, \bibinfo {author} {\bibfnamefont {C.~D.}\ \bibnamefont {Wright}},
  \bibinfo {author} {\bibfnamefont {A.}~\bibnamefont {Sebastian}}, \bibinfo
  {author} {\bibfnamefont {T.~J.}\ \bibnamefont {Kippenberg}}, \bibinfo
  {author} {\bibfnamefont {W.~H.~P.}\ \bibnamefont {Pernice}}, \ and\ \bibinfo
  {author} {\bibfnamefont {H.}~\bibnamefont {Bhaskaran}},\ }\bibfield  {title}
  {\enquote {\bibinfo {title} {Parallel convolutional processing using an
  integrated photonic tensor core},}\ }\href {\doibase
  10.1038/s41586-020-03070-1} {\bibfield  {journal} {\bibinfo  {journal}
  {Nature}\ }\textbf {\bibinfo {volume} {589}},\ \bibinfo {pages} {52}
  (\bibinfo {year} {2021})}\BibitemShut {NoStop}%
\bibitem [{\citenamefont {Xu}\ \emph {et~al.}(2021{\natexlab{a}})\citenamefont
  {Xu}, \citenamefont {Tan}, \citenamefont {Corcoran}, \citenamefont {Wu},
  \citenamefont {Boes}, \citenamefont {Nguyen}, \citenamefont {Chu},
  \citenamefont {Little}, \citenamefont {Hicks}, \citenamefont {Morandotti},
  \citenamefont {Mitchell},\ and\ \citenamefont {Moss}}]{Xu:21}%
  \BibitemOpen
  \bibfield  {author} {\bibinfo {author} {\bibfnamefont {X.}~\bibnamefont
  {Xu}}, \bibinfo {author} {\bibfnamefont {M.}~\bibnamefont {Tan}}, \bibinfo
  {author} {\bibfnamefont {B.}~\bibnamefont {Corcoran}}, \bibinfo {author}
  {\bibfnamefont {J.}~\bibnamefont {Wu}}, \bibinfo {author} {\bibfnamefont
  {A.}~\bibnamefont {Boes}}, \bibinfo {author} {\bibfnamefont {T.~G.}\
  \bibnamefont {Nguyen}}, \bibinfo {author} {\bibfnamefont {S.~T.}\
  \bibnamefont {Chu}}, \bibinfo {author} {\bibfnamefont {B.~E.}\ \bibnamefont
  {Little}}, \bibinfo {author} {\bibfnamefont {D.~G.}\ \bibnamefont {Hicks}},
  \bibinfo {author} {\bibfnamefont {R.}~\bibnamefont {Morandotti}}, \bibinfo
  {author} {\bibfnamefont {A.}~\bibnamefont {Mitchell}}, \ and\ \bibinfo
  {author} {\bibfnamefont {D.~J.}\ \bibnamefont {Moss}},\ }\bibfield  {title}
  {\enquote {\bibinfo {title} {11 tops photonic convolutional accelerator for
  optical neural networks},}\ }\href {\doibase 10.1038/s41586-020-03063-0}
  {\bibfield  {journal} {\bibinfo  {journal} {Nature}\ }\textbf {\bibinfo
  {volume} {589}},\ \bibinfo {pages} {44} (\bibinfo {year}
  {2021}{\natexlab{a}})}\BibitemShut {NoStop}%
\bibitem [{\citenamefont {Liang}\ \emph {et~al.}(2015)\citenamefont {Liang},
  \citenamefont {Eliyahu}, \citenamefont {Ilchenko}, \citenamefont
  {Savchenkov}, \citenamefont {Matsko}, \citenamefont {Seidel},\ and\
  \citenamefont {Maleki}}]{Liang:15}%
  \BibitemOpen
  \bibfield  {author} {\bibinfo {author} {\bibfnamefont {W.}~\bibnamefont
  {Liang}}, \bibinfo {author} {\bibfnamefont {D.}~\bibnamefont {Eliyahu}},
  \bibinfo {author} {\bibfnamefont {V.~S.}\ \bibnamefont {Ilchenko}}, \bibinfo
  {author} {\bibfnamefont {A.~A.}\ \bibnamefont {Savchenkov}}, \bibinfo
  {author} {\bibfnamefont {A.~B.}\ \bibnamefont {Matsko}}, \bibinfo {author}
  {\bibfnamefont {D.}~\bibnamefont {Seidel}}, \ and\ \bibinfo {author}
  {\bibfnamefont {L.}~\bibnamefont {Maleki}},\ }\bibfield  {title} {\enquote
  {\bibinfo {title} {High spectral purity kerr frequency comb radio frequency
  photonic oscillator},}\ }\href {https://doi.org/10.1038/ncomms8957}
  {\bibfield  {journal} {\bibinfo  {journal} {Nature Communications}\ }\textbf
  {\bibinfo {volume} {6}},\ \bibinfo {pages} {7957} (\bibinfo {year}
  {2015})}\BibitemShut {NoStop}%
\bibitem [{\citenamefont {Liu}\ \emph {et~al.}(2020)\citenamefont {Liu},
  \citenamefont {Lucas}, \citenamefont {Raja}, \citenamefont {He},
  \citenamefont {Riemensberger}, \citenamefont {Wang}, \citenamefont {Karpov},
  \citenamefont {Guo}, \citenamefont {Bouchand},\ and\ \citenamefont
  {Kippenberg}}]{Liu:20}%
  \BibitemOpen
  \bibfield  {author} {\bibinfo {author} {\bibfnamefont {J.}~\bibnamefont
  {Liu}}, \bibinfo {author} {\bibfnamefont {E.}~\bibnamefont {Lucas}}, \bibinfo
  {author} {\bibfnamefont {A.~S.}\ \bibnamefont {Raja}}, \bibinfo {author}
  {\bibfnamefont {J.}~\bibnamefont {He}}, \bibinfo {author} {\bibfnamefont
  {J.}~\bibnamefont {Riemensberger}}, \bibinfo {author} {\bibfnamefont {R.~N.}\
  \bibnamefont {Wang}}, \bibinfo {author} {\bibfnamefont {M.}~\bibnamefont
  {Karpov}}, \bibinfo {author} {\bibfnamefont {H.}~\bibnamefont {Guo}},
  \bibinfo {author} {\bibfnamefont {R.}~\bibnamefont {Bouchand}}, \ and\
  \bibinfo {author} {\bibfnamefont {T.~J.}\ \bibnamefont {Kippenberg}},\
  }\bibfield  {title} {\enquote {\bibinfo {title} {Photonic microwave
  generation in the x- and k-band using integrated soliton microcombs},}\
  }\href {\doibase 10.1038/s41566-020-0617-x} {\bibfield  {journal} {\bibinfo
  {journal} {Nature Photonics}\ }\textbf {\bibinfo {volume} {14}},\ \bibinfo
  {pages} {486} (\bibinfo {year} {2020})}\BibitemShut {NoStop}%
\bibitem [{\citenamefont {Yang}\ \emph {et~al.}(2019)\citenamefont {Yang},
  \citenamefont {Shen}, \citenamefont {Wang}, \citenamefont {Tran},
  \citenamefont {Zhang}, \citenamefont {Yang}, \citenamefont {Wu},
  \citenamefont {Bao}, \citenamefont {Bowers}, \citenamefont {Yariv},\ and\
  \citenamefont {Vahala}}]{Yang:19}%
  \BibitemOpen
  \bibfield  {author} {\bibinfo {author} {\bibfnamefont {Q.-F.}\ \bibnamefont
  {Yang}}, \bibinfo {author} {\bibfnamefont {B.}~\bibnamefont {Shen}}, \bibinfo
  {author} {\bibfnamefont {H.}~\bibnamefont {Wang}}, \bibinfo {author}
  {\bibfnamefont {M.}~\bibnamefont {Tran}}, \bibinfo {author} {\bibfnamefont
  {Z.}~\bibnamefont {Zhang}}, \bibinfo {author} {\bibfnamefont {K.~Y.}\
  \bibnamefont {Yang}}, \bibinfo {author} {\bibfnamefont {L.}~\bibnamefont
  {Wu}}, \bibinfo {author} {\bibfnamefont {C.}~\bibnamefont {Bao}}, \bibinfo
  {author} {\bibfnamefont {J.}~\bibnamefont {Bowers}}, \bibinfo {author}
  {\bibfnamefont {A.}~\bibnamefont {Yariv}}, \ and\ \bibinfo {author}
  {\bibfnamefont {K.}~\bibnamefont {Vahala}},\ }\bibfield  {title} {\enquote
  {\bibinfo {title} {Vernier spectrometer using counterpropagating soliton
  microcombs},}\ }\href {http://doi.org/10.5281/zenodo.2542265} {\bibfield
  {journal} {\bibinfo  {journal} {Science}\ }\textbf {\bibinfo {volume}
  {363}},\ \bibinfo {pages} {965} (\bibinfo {year} {2019})}\BibitemShut
  {NoStop}%
\bibitem [{\citenamefont {Avik}\ \emph {et~al.}()\citenamefont {Avik},
  \citenamefont {Chaitanya}, \citenamefont {Xingchen}, \citenamefont {Jaime},
  \citenamefont {Yoshitomo}, \citenamefont {Kevin}, \citenamefont {L.},\ and\
  \citenamefont {Michal}}]{Dutt:18}%
  \BibitemOpen
  \bibfield  {author} {\bibinfo {author} {\bibfnamefont {D.}~\bibnamefont
  {Avik}}, \bibinfo {author} {\bibfnamefont {J.}~\bibnamefont {Chaitanya}},
  \bibinfo {author} {\bibfnamefont {J.}~\bibnamefont {Xingchen}}, \bibinfo
  {author} {\bibfnamefont {C.}~\bibnamefont {Jaime}}, \bibinfo {author}
  {\bibfnamefont {O.}~\bibnamefont {Yoshitomo}}, \bibinfo {author}
  {\bibfnamefont {L.}~\bibnamefont {Kevin}}, \bibinfo {author} {\bibfnamefont
  {G.~A.}\ \bibnamefont {L.}}, \ and\ \bibinfo {author} {\bibfnamefont
  {L.}~\bibnamefont {Michal}},\ }\bibfield  {title} {\enquote {\bibinfo {title}
  {On-chip dual-comb source for spectroscopy},}\ }\href {\doibase
  10.1126/sciadv.1701858} {\bibfield  {journal} {\bibinfo  {journal} {Science
  Advances}\ }\textbf {\bibinfo {volume} {4}},\ \bibinfo {pages}
  {e1701858}}\BibitemShut {NoStop}%
\bibitem [{\citenamefont {Wang}\ \emph {et~al.}(2020)\citenamefont {Wang},
  \citenamefont {Wang}, \citenamefont {Niu}, \citenamefont {Wang},
  \citenamefont {Zou}, \citenamefont {Dong}, \citenamefont {Little},
  \citenamefont {Chu}, \citenamefont {Liu}, \citenamefont {Hao}, \citenamefont
  {Liu}, \citenamefont {Wang}, \citenamefont {Yin}, \citenamefont {He},
  \citenamefont {Zhang}, \citenamefont {Zhao}, \citenamefont {Han},
  \citenamefont {Guo},\ and\ \citenamefont {Chen}}]{WangFX:20}%
  \BibitemOpen
  \bibfield  {author} {\bibinfo {author} {\bibfnamefont {F.-X.}\ \bibnamefont
  {Wang}}, \bibinfo {author} {\bibfnamefont {W.}~\bibnamefont {Wang}}, \bibinfo
  {author} {\bibfnamefont {R.}~\bibnamefont {Niu}}, \bibinfo {author}
  {\bibfnamefont {X.}~\bibnamefont {Wang}}, \bibinfo {author} {\bibfnamefont
  {C.-L.}\ \bibnamefont {Zou}}, \bibinfo {author} {\bibfnamefont {C.-H.}\
  \bibnamefont {Dong}}, \bibinfo {author} {\bibfnamefont {B.~E.}\ \bibnamefont
  {Little}}, \bibinfo {author} {\bibfnamefont {S.~T.}\ \bibnamefont {Chu}},
  \bibinfo {author} {\bibfnamefont {H.}~\bibnamefont {Liu}}, \bibinfo {author}
  {\bibfnamefont {P.}~\bibnamefont {Hao}}, \bibinfo {author} {\bibfnamefont
  {S.}~\bibnamefont {Liu}}, \bibinfo {author} {\bibfnamefont {S.}~\bibnamefont
  {Wang}}, \bibinfo {author} {\bibfnamefont {Z.-Q.}\ \bibnamefont {Yin}},
  \bibinfo {author} {\bibfnamefont {D.-Y.}\ \bibnamefont {He}}, \bibinfo
  {author} {\bibfnamefont {W.}~\bibnamefont {Zhang}}, \bibinfo {author}
  {\bibfnamefont {W.}~\bibnamefont {Zhao}}, \bibinfo {author} {\bibfnamefont
  {Z.-F.}\ \bibnamefont {Han}}, \bibinfo {author} {\bibfnamefont {G.-C.}\
  \bibnamefont {Guo}}, \ and\ \bibinfo {author} {\bibfnamefont
  {W.}~\bibnamefont {Chen}},\ }\bibfield  {title} {\enquote {\bibinfo {title}
  {Quantum key distribution with on-chip dissipative kerr soliton},}\ }\href
  {\doibase https://doi.org/10.1002/lpor.201900190} {\bibfield  {journal}
  {\bibinfo  {journal} {Laser \& Photonics Reviews}\ }\textbf {\bibinfo
  {volume} {14}},\ \bibinfo {pages} {1900190} (\bibinfo {year}
  {2020})}\BibitemShut {NoStop}%
\bibitem [{\citenamefont {Spencer}\ \emph {et~al.}(2018)\citenamefont
  {Spencer}, \citenamefont {Drake}, \citenamefont {Briles}, \citenamefont
  {Stone}, \citenamefont {Sinclair}, \citenamefont {Fredrick}, \citenamefont
  {Li}, \citenamefont {Westly}, \citenamefont {Ilic}, \citenamefont
  {Bluestone}, \citenamefont {Volet}, \citenamefont {Komljenovic},
  \citenamefont {Chang}, \citenamefont {Lee}, \citenamefont {Oh}, \citenamefont
  {Suh}, \citenamefont {Yang}, \citenamefont {Pfeiffer}, \citenamefont
  {Kippenberg}, \citenamefont {Norberg}, \citenamefont {Theogarajan},
  \citenamefont {Vahala}, \citenamefont {Newbury}, \citenamefont {Srinivasan},
  \citenamefont {Bowers}, \citenamefont {Diddams},\ and\ \citenamefont
  {Papp}}]{Spencer:18}%
  \BibitemOpen
  \bibfield  {author} {\bibinfo {author} {\bibfnamefont {D.~T.}\ \bibnamefont
  {Spencer}}, \bibinfo {author} {\bibfnamefont {T.}~\bibnamefont {Drake}},
  \bibinfo {author} {\bibfnamefont {T.~C.}\ \bibnamefont {Briles}}, \bibinfo
  {author} {\bibfnamefont {J.}~\bibnamefont {Stone}}, \bibinfo {author}
  {\bibfnamefont {L.~C.}\ \bibnamefont {Sinclair}}, \bibinfo {author}
  {\bibfnamefont {C.}~\bibnamefont {Fredrick}}, \bibinfo {author}
  {\bibfnamefont {Q.}~\bibnamefont {Li}}, \bibinfo {author} {\bibfnamefont
  {D.}~\bibnamefont {Westly}}, \bibinfo {author} {\bibfnamefont {B.~R.}\
  \bibnamefont {Ilic}}, \bibinfo {author} {\bibfnamefont {A.}~\bibnamefont
  {Bluestone}}, \bibinfo {author} {\bibfnamefont {N.}~\bibnamefont {Volet}},
  \bibinfo {author} {\bibfnamefont {T.}~\bibnamefont {Komljenovic}}, \bibinfo
  {author} {\bibfnamefont {L.}~\bibnamefont {Chang}}, \bibinfo {author}
  {\bibfnamefont {S.~H.}\ \bibnamefont {Lee}}, \bibinfo {author} {\bibfnamefont
  {D.~Y.}\ \bibnamefont {Oh}}, \bibinfo {author} {\bibfnamefont {M.-G.}\
  \bibnamefont {Suh}}, \bibinfo {author} {\bibfnamefont {K.~Y.}\ \bibnamefont
  {Yang}}, \bibinfo {author} {\bibfnamefont {M.~H.~P.}\ \bibnamefont
  {Pfeiffer}}, \bibinfo {author} {\bibfnamefont {T.~J.}\ \bibnamefont
  {Kippenberg}}, \bibinfo {author} {\bibfnamefont {E.}~\bibnamefont {Norberg}},
  \bibinfo {author} {\bibfnamefont {L.}~\bibnamefont {Theogarajan}}, \bibinfo
  {author} {\bibfnamefont {K.}~\bibnamefont {Vahala}}, \bibinfo {author}
  {\bibfnamefont {N.~R.}\ \bibnamefont {Newbury}}, \bibinfo {author}
  {\bibfnamefont {K.}~\bibnamefont {Srinivasan}}, \bibinfo {author}
  {\bibfnamefont {J.~E.}\ \bibnamefont {Bowers}}, \bibinfo {author}
  {\bibfnamefont {S.~A.}\ \bibnamefont {Diddams}}, \ and\ \bibinfo {author}
  {\bibfnamefont {S.~B.}\ \bibnamefont {Papp}},\ }\bibfield  {title} {\enquote
  {\bibinfo {title} {An optical-frequency synthesizer using integrated
  photonics},}\ }\href {\doibase 10.1038/s41586-018-0065-7} {\bibfield
  {journal} {\bibinfo  {journal} {Nature}\ }\textbf {\bibinfo {volume} {557}},\
  \bibinfo {pages} {81} (\bibinfo {year} {2018})}\BibitemShut {NoStop}%
\bibitem [{\citenamefont {Newman}\ \emph {et~al.}(2019)\citenamefont {Newman},
  \citenamefont {Maurice}, \citenamefont {Drake}, \citenamefont {Stone},
  \citenamefont {Briles}, \citenamefont {Spencer}, \citenamefont {Fredrick},
  \citenamefont {Li}, \citenamefont {Westly}, \citenamefont {Ilic},
  \citenamefont {Shen}, \citenamefont {Suh}, \citenamefont {Yang},
  \citenamefont {Johnson}, \citenamefont {Johnson}, \citenamefont {Hollberg},
  \citenamefont {Vahala}, \citenamefont {Srinivasan}, \citenamefont {Diddams},
  \citenamefont {Kitching}, \citenamefont {Papp},\ and\ \citenamefont
  {Hummon}}]{Newman:19}%
  \BibitemOpen
  \bibfield  {author} {\bibinfo {author} {\bibfnamefont {Z.~L.}\ \bibnamefont
  {Newman}}, \bibinfo {author} {\bibfnamefont {V.}~\bibnamefont {Maurice}},
  \bibinfo {author} {\bibfnamefont {T.}~\bibnamefont {Drake}}, \bibinfo
  {author} {\bibfnamefont {J.~R.}\ \bibnamefont {Stone}}, \bibinfo {author}
  {\bibfnamefont {T.~C.}\ \bibnamefont {Briles}}, \bibinfo {author}
  {\bibfnamefont {D.~T.}\ \bibnamefont {Spencer}}, \bibinfo {author}
  {\bibfnamefont {C.}~\bibnamefont {Fredrick}}, \bibinfo {author}
  {\bibfnamefont {Q.}~\bibnamefont {Li}}, \bibinfo {author} {\bibfnamefont
  {D.}~\bibnamefont {Westly}}, \bibinfo {author} {\bibfnamefont {B.~R.}\
  \bibnamefont {Ilic}}, \bibinfo {author} {\bibfnamefont {B.}~\bibnamefont
  {Shen}}, \bibinfo {author} {\bibfnamefont {M.-G.}\ \bibnamefont {Suh}},
  \bibinfo {author} {\bibfnamefont {K.~Y.}\ \bibnamefont {Yang}}, \bibinfo
  {author} {\bibfnamefont {C.}~\bibnamefont {Johnson}}, \bibinfo {author}
  {\bibfnamefont {D.~M.~S.}\ \bibnamefont {Johnson}}, \bibinfo {author}
  {\bibfnamefont {L.}~\bibnamefont {Hollberg}}, \bibinfo {author}
  {\bibfnamefont {K.~J.}\ \bibnamefont {Vahala}}, \bibinfo {author}
  {\bibfnamefont {K.}~\bibnamefont {Srinivasan}}, \bibinfo {author}
  {\bibfnamefont {S.~A.}\ \bibnamefont {Diddams}}, \bibinfo {author}
  {\bibfnamefont {J.}~\bibnamefont {Kitching}}, \bibinfo {author}
  {\bibfnamefont {S.~B.}\ \bibnamefont {Papp}}, \ and\ \bibinfo {author}
  {\bibfnamefont {M.~T.}\ \bibnamefont {Hummon}},\ }\bibfield  {title}
  {\enquote {\bibinfo {title} {Architecture for the photonic integration of an
  optical atomic clock},}\ }\href {\doibase 10.1364/OPTICA.6.000680} {\bibfield
   {journal} {\bibinfo  {journal} {Optica}\ }\textbf {\bibinfo {volume} {6}},\
  \bibinfo {pages} {680} (\bibinfo {year} {2019})}\BibitemShut {NoStop}%
\bibitem [{\citenamefont {Jost}\ \emph {et~al.}(2015)\citenamefont {Jost},
  \citenamefont {Herr}, \citenamefont {Lecaplain}, \citenamefont {Brasch},
  \citenamefont {Pfeiffer},\ and\ \citenamefont {Kippenberg}}]{Jost:15}%
  \BibitemOpen
  \bibfield  {author} {\bibinfo {author} {\bibfnamefont {J.~D.}\ \bibnamefont
  {Jost}}, \bibinfo {author} {\bibfnamefont {T.}~\bibnamefont {Herr}}, \bibinfo
  {author} {\bibfnamefont {C.}~\bibnamefont {Lecaplain}}, \bibinfo {author}
  {\bibfnamefont {V.}~\bibnamefont {Brasch}}, \bibinfo {author} {\bibfnamefont
  {M.~H.~P.}\ \bibnamefont {Pfeiffer}}, \ and\ \bibinfo {author} {\bibfnamefont
  {T.~J.}\ \bibnamefont {Kippenberg}},\ }\bibfield  {title} {\enquote {\bibinfo
  {title} {Counting the cycles of light using a self-referenced optical
  microresonator},}\ }\href {\doibase 10.1364/OPTICA.2.000706} {\bibfield
  {journal} {\bibinfo  {journal} {Optica}\ }\textbf {\bibinfo {volume} {2}},\
  \bibinfo {pages} {706} (\bibinfo {year} {2015})}\BibitemShut {NoStop}%
\bibitem [{\citenamefont {Del'Haye}\ \emph {et~al.}(2016)\citenamefont
  {Del'Haye}, \citenamefont {Coillet}, \citenamefont {Fortier}, \citenamefont
  {Beha}, \citenamefont {Cole}, \citenamefont {Yang}, \citenamefont {Lee},
  \citenamefont {Vahala}, \citenamefont {Papp},\ and\ \citenamefont
  {Diddams}}]{DelHaye:16}%
  \BibitemOpen
  \bibfield  {author} {\bibinfo {author} {\bibfnamefont {P.}~\bibnamefont
  {Del'Haye}}, \bibinfo {author} {\bibfnamefont {A.}~\bibnamefont {Coillet}},
  \bibinfo {author} {\bibfnamefont {T.}~\bibnamefont {Fortier}}, \bibinfo
  {author} {\bibfnamefont {K.}~\bibnamefont {Beha}}, \bibinfo {author}
  {\bibfnamefont {D.~C.}\ \bibnamefont {Cole}}, \bibinfo {author}
  {\bibfnamefont {K.~Y.}\ \bibnamefont {Yang}}, \bibinfo {author}
  {\bibfnamefont {H.}~\bibnamefont {Lee}}, \bibinfo {author} {\bibfnamefont
  {K.~J.}\ \bibnamefont {Vahala}}, \bibinfo {author} {\bibfnamefont {S.~B.}\
  \bibnamefont {Papp}}, \ and\ \bibinfo {author} {\bibfnamefont {S.~A.}\
  \bibnamefont {Diddams}},\ }\bibfield  {title} {\enquote {\bibinfo {title}
  {Phase-coherent microwave-to-optical link with a self-referenced
  microcomb},}\ }\href {https://doi.org/10.1038/nphoton.2016.105} {\bibfield
  {journal} {\bibinfo  {journal} {Nature Photonics}\ }\textbf {\bibinfo
  {volume} {10}},\ \bibinfo {pages} {516} (\bibinfo {year} {2016})}\BibitemShut
  {NoStop}%
\bibitem [{\citenamefont {Brasch}\ \emph {et~al.}(2017)\citenamefont {Brasch},
  \citenamefont {Lucas}, \citenamefont {Jost}, \citenamefont {Geiselmann},\
  and\ \citenamefont {Kippenberg}}]{Brasch:17}%
  \BibitemOpen
  \bibfield  {author} {\bibinfo {author} {\bibfnamefont {V.}~\bibnamefont
  {Brasch}}, \bibinfo {author} {\bibfnamefont {E.}~\bibnamefont {Lucas}},
  \bibinfo {author} {\bibfnamefont {J.~D.}\ \bibnamefont {Jost}}, \bibinfo
  {author} {\bibfnamefont {M.}~\bibnamefont {Geiselmann}}, \ and\ \bibinfo
  {author} {\bibfnamefont {T.~J.}\ \bibnamefont {Kippenberg}},\ }\bibfield
  {title} {\enquote {\bibinfo {title} {Self-referenced photonic chip soliton
  kerr frequency comb},}\ }\href {\doibase 10.1038/lsa.2016.202} {\bibfield
  {journal} {\bibinfo  {journal} {Light: Science \& Applications}\ }\textbf
  {\bibinfo {volume} {6}},\ \bibinfo {pages} {e16202} (\bibinfo {year}
  {2017})}\BibitemShut {NoStop}%
\bibitem [{\citenamefont {Brasch}\ \emph
  {et~al.}(2016{\natexlab{a}})\citenamefont {Brasch}, \citenamefont
  {Geiselmann}, \citenamefont {Herr}, \citenamefont {Lihachev}, \citenamefont
  {Pfeiffer}, \citenamefont {Gorodetsky},\ and\ \citenamefont
  {Kippenberg}}]{Brasch:15}%
  \BibitemOpen
  \bibfield  {author} {\bibinfo {author} {\bibfnamefont {V.}~\bibnamefont
  {Brasch}}, \bibinfo {author} {\bibfnamefont {M.}~\bibnamefont {Geiselmann}},
  \bibinfo {author} {\bibfnamefont {T.}~\bibnamefont {Herr}}, \bibinfo {author}
  {\bibfnamefont {G.}~\bibnamefont {Lihachev}}, \bibinfo {author}
  {\bibfnamefont {M.~H.~P.}\ \bibnamefont {Pfeiffer}}, \bibinfo {author}
  {\bibfnamefont {M.~L.}\ \bibnamefont {Gorodetsky}}, \ and\ \bibinfo {author}
  {\bibfnamefont {T.~J.}\ \bibnamefont {Kippenberg}},\ }\bibfield  {title}
  {\enquote {\bibinfo {title} {Photonic chip{\textendash}based optical
  frequency comb using soliton cherenkov radiation},}\ }\href {\doibase
  10.1126/science.aad4811} {\bibfield  {journal} {\bibinfo  {journal}
  {Science}\ }\textbf {\bibinfo {volume} {351}},\ \bibinfo {pages} {357}
  (\bibinfo {year} {2016}{\natexlab{a}})}\BibitemShut {NoStop}%
\bibitem [{\citenamefont {Yu}\ \emph {et~al.}(2019)\citenamefont {Yu},
  \citenamefont {Briles}, \citenamefont {Moille}, \citenamefont {Lu},
  \citenamefont {Diddams}, \citenamefont {Srinivasan},\ and\ \citenamefont
  {Papp}}]{Yu:19}%
  \BibitemOpen
  \bibfield  {author} {\bibinfo {author} {\bibfnamefont {S.-P.}\ \bibnamefont
  {Yu}}, \bibinfo {author} {\bibfnamefont {T.~C.}\ \bibnamefont {Briles}},
  \bibinfo {author} {\bibfnamefont {G.~T.}\ \bibnamefont {Moille}}, \bibinfo
  {author} {\bibfnamefont {X.}~\bibnamefont {Lu}}, \bibinfo {author}
  {\bibfnamefont {S.~A.}\ \bibnamefont {Diddams}}, \bibinfo {author}
  {\bibfnamefont {K.}~\bibnamefont {Srinivasan}}, \ and\ \bibinfo {author}
  {\bibfnamefont {S.~B.}\ \bibnamefont {Papp}},\ }\bibfield  {title} {\enquote
  {\bibinfo {title} {Tuning kerr-soliton frequency combs to atomic
  resonances},}\ }\href {\doibase 10.1103/PhysRevApplied.11.044017} {\bibfield
  {journal} {\bibinfo  {journal} {Phys. Rev. Applied}\ }\textbf {\bibinfo
  {volume} {11}},\ \bibinfo {pages} {044017} (\bibinfo {year}
  {2019})}\BibitemShut {NoStop}%
\bibitem [{\citenamefont {Liu}\ \emph {et~al.}(2021)\citenamefont {Liu},
  \citenamefont {Gong}, \citenamefont {Bruch}, \citenamefont {Surya},
  \citenamefont {Lu},\ and\ \citenamefont {Tang}}]{LiuX:21}%
  \BibitemOpen
  \bibfield  {author} {\bibinfo {author} {\bibfnamefont {X.}~\bibnamefont
  {Liu}}, \bibinfo {author} {\bibfnamefont {Z.}~\bibnamefont {Gong}}, \bibinfo
  {author} {\bibfnamefont {A.~W.}\ \bibnamefont {Bruch}}, \bibinfo {author}
  {\bibfnamefont {J.~B.}\ \bibnamefont {Surya}}, \bibinfo {author}
  {\bibfnamefont {J.}~\bibnamefont {Lu}}, \ and\ \bibinfo {author}
  {\bibfnamefont {H.~X.}\ \bibnamefont {Tang}},\ }\bibfield  {title} {\enquote
  {\bibinfo {title} {Aluminum nitride nanophotonics for beyond-octave soliton
  microcomb generation and self-referencing},}\ }\href {\doibase
  10.1038/s41467-021-25751-9} {\bibfield  {journal} {\bibinfo  {journal}
  {Nature Communications}\ }\textbf {\bibinfo {volume} {12}},\ \bibinfo {pages}
  {5428} (\bibinfo {year} {2021})}\BibitemShut {NoStop}%
\bibitem [{\citenamefont {Gong}\ \emph {et~al.}(2020)\citenamefont {Gong},
  \citenamefont {Liu}, \citenamefont {Xu},\ and\ \citenamefont
  {Tang}}]{Gong:20}%
  \BibitemOpen
  \bibfield  {author} {\bibinfo {author} {\bibfnamefont {Z.}~\bibnamefont
  {Gong}}, \bibinfo {author} {\bibfnamefont {X.}~\bibnamefont {Liu}}, \bibinfo
  {author} {\bibfnamefont {Y.}~\bibnamefont {Xu}}, \ and\ \bibinfo {author}
  {\bibfnamefont {H.~X.}\ \bibnamefont {Tang}},\ }\bibfield  {title} {\enquote
  {\bibinfo {title} {Near-octave lithium niobate soliton microcomb},}\ }\href
  {\doibase 10.1364/OPTICA.400994} {\bibfield  {journal} {\bibinfo  {journal}
  {Optica}\ }\textbf {\bibinfo {volume} {7}},\ \bibinfo {pages} {1275}
  (\bibinfo {year} {2020})}\BibitemShut {NoStop}%
\bibitem [{\citenamefont {Brasch}\ \emph
  {et~al.}(2016{\natexlab{b}})\citenamefont {Brasch}, \citenamefont
  {Geiselmann}, \citenamefont {Pfeiffer},\ and\ \citenamefont
  {Kippenberg}}]{Brasch:16}%
  \BibitemOpen
  \bibfield  {author} {\bibinfo {author} {\bibfnamefont {V.}~\bibnamefont
  {Brasch}}, \bibinfo {author} {\bibfnamefont {M.}~\bibnamefont {Geiselmann}},
  \bibinfo {author} {\bibfnamefont {M.~H.~P.}\ \bibnamefont {Pfeiffer}}, \ and\
  \bibinfo {author} {\bibfnamefont {T.~J.}\ \bibnamefont {Kippenberg}},\
  }\bibfield  {title} {\enquote {\bibinfo {title} {Bringing short-lived
  dissipative kerr soliton states in microresonators into a steady state},}\
  }\href {\doibase 10.1364/OE.24.029312} {\bibfield  {journal} {\bibinfo
  {journal} {Opt. Express}\ }\textbf {\bibinfo {volume} {24}},\ \bibinfo
  {pages} {29312} (\bibinfo {year} {2016}{\natexlab{b}})}\BibitemShut {NoStop}%
\bibitem [{\citenamefont {Yi}\ \emph {et~al.}(2016)\citenamefont {Yi},
  \citenamefont {Yang}, \citenamefont {Yang},\ and\ \citenamefont
  {Vahala}}]{Yi:16b}%
  \BibitemOpen
  \bibfield  {author} {\bibinfo {author} {\bibfnamefont {X.}~\bibnamefont
  {Yi}}, \bibinfo {author} {\bibfnamefont {Q.-F.}\ \bibnamefont {Yang}},
  \bibinfo {author} {\bibfnamefont {K.~Y.}\ \bibnamefont {Yang}}, \ and\
  \bibinfo {author} {\bibfnamefont {K.}~\bibnamefont {Vahala}},\ }\bibfield
  {title} {\enquote {\bibinfo {title} {Active capture and stabilization of
  temporal solitons in microresonators},}\ }\href {\doibase
  10.1364/OL.41.002037} {\bibfield  {journal} {\bibinfo  {journal} {Opt.
  Lett.}\ }\textbf {\bibinfo {volume} {41}},\ \bibinfo {pages} {2037} (\bibinfo
  {year} {2016})}\BibitemShut {NoStop}%
\bibitem [{\citenamefont {Stone}\ \emph {et~al.}(2018)\citenamefont {Stone},
  \citenamefont {Briles}, \citenamefont {Drake}, \citenamefont {Spencer},
  \citenamefont {Carlson}, \citenamefont {Diddams},\ and\ \citenamefont
  {Papp}}]{Stone:18}%
  \BibitemOpen
  \bibfield  {author} {\bibinfo {author} {\bibfnamefont {J.~R.}\ \bibnamefont
  {Stone}}, \bibinfo {author} {\bibfnamefont {T.~C.}\ \bibnamefont {Briles}},
  \bibinfo {author} {\bibfnamefont {T.~E.}\ \bibnamefont {Drake}}, \bibinfo
  {author} {\bibfnamefont {D.~T.}\ \bibnamefont {Spencer}}, \bibinfo {author}
  {\bibfnamefont {D.~R.}\ \bibnamefont {Carlson}}, \bibinfo {author}
  {\bibfnamefont {S.~A.}\ \bibnamefont {Diddams}}, \ and\ \bibinfo {author}
  {\bibfnamefont {S.~B.}\ \bibnamefont {Papp}},\ }\bibfield  {title} {\enquote
  {\bibinfo {title} {Thermal and nonlinear dissipative-soliton dynamics in
  kerr-microresonator frequency combs},}\ }\href {\doibase
  10.1103/PhysRevLett.121.063902} {\bibfield  {journal} {\bibinfo  {journal}
  {Phys. Rev. Lett.}\ }\textbf {\bibinfo {volume} {121}},\ \bibinfo {pages}
  {063902} (\bibinfo {year} {2018})}\BibitemShut {NoStop}%
\bibitem [{\citenamefont {Zhou}\ \emph {et~al.}(2019)\citenamefont {Zhou},
  \citenamefont {Geng}, \citenamefont {Cui}, \citenamefont {Huang},
  \citenamefont {Zhou}, \citenamefont {Qiu},\ and\ \citenamefont
  {Wei~Wong}}]{Zhou:19}%
  \BibitemOpen
  \bibfield  {author} {\bibinfo {author} {\bibfnamefont {H.}~\bibnamefont
  {Zhou}}, \bibinfo {author} {\bibfnamefont {Y.}~\bibnamefont {Geng}}, \bibinfo
  {author} {\bibfnamefont {W.}~\bibnamefont {Cui}}, \bibinfo {author}
  {\bibfnamefont {S.-W.}\ \bibnamefont {Huang}}, \bibinfo {author}
  {\bibfnamefont {Q.}~\bibnamefont {Zhou}}, \bibinfo {author} {\bibfnamefont
  {K.}~\bibnamefont {Qiu}}, \ and\ \bibinfo {author} {\bibfnamefont
  {C.}~\bibnamefont {Wei~Wong}},\ }\bibfield  {title} {\enquote {\bibinfo
  {title} {Soliton bursts and deterministic dissipative kerr soliton generation
  in auxiliary-assisted microcavities},}\ }\href {\doibase
  10.1038/s41377-019-0161-y} {\bibfield  {journal} {\bibinfo  {journal} {Light:
  Science \& Applications}\ }\textbf {\bibinfo {volume} {8}},\ \bibinfo {pages}
  {50} (\bibinfo {year} {2019})}\BibitemShut {NoStop}%
\bibitem [{\citenamefont {Lu}\ \emph {et~al.}(2021)\citenamefont {Lu},
  \citenamefont {Chen}, \citenamefont {Wang}, \citenamefont {Yao},
  \citenamefont {Wang}, \citenamefont {Yu}, \citenamefont {Little},
  \citenamefont {Chu}, \citenamefont {Gong}, \citenamefont {Zhao},
  \citenamefont {Yi}, \citenamefont {Xiao},\ and\ \citenamefont
  {Zhang}}]{LuZ:21}%
  \BibitemOpen
  \bibfield  {author} {\bibinfo {author} {\bibfnamefont {Z.}~\bibnamefont
  {Lu}}, \bibinfo {author} {\bibfnamefont {H.-J.}\ \bibnamefont {Chen}},
  \bibinfo {author} {\bibfnamefont {W.}~\bibnamefont {Wang}}, \bibinfo {author}
  {\bibfnamefont {L.}~\bibnamefont {Yao}}, \bibinfo {author} {\bibfnamefont
  {Y.}~\bibnamefont {Wang}}, \bibinfo {author} {\bibfnamefont {Y.}~\bibnamefont
  {Yu}}, \bibinfo {author} {\bibfnamefont {B.~E.}\ \bibnamefont {Little}},
  \bibinfo {author} {\bibfnamefont {S.~T.}\ \bibnamefont {Chu}}, \bibinfo
  {author} {\bibfnamefont {Q.}~\bibnamefont {Gong}}, \bibinfo {author}
  {\bibfnamefont {W.}~\bibnamefont {Zhao}}, \bibinfo {author} {\bibfnamefont
  {X.}~\bibnamefont {Yi}}, \bibinfo {author} {\bibfnamefont {Y.-F.}\
  \bibnamefont {Xiao}}, \ and\ \bibinfo {author} {\bibfnamefont
  {W.}~\bibnamefont {Zhang}},\ }\bibfield  {title} {\enquote {\bibinfo {title}
  {Synthesized soliton crystals},}\ }\href {\doibase
  10.1038/s41467-021-23172-2} {\bibfield  {journal} {\bibinfo  {journal}
  {Nature Communications}\ }\textbf {\bibinfo {volume} {12}},\ \bibinfo {pages}
  {3179} (\bibinfo {year} {2021})}\BibitemShut {NoStop}%
\bibitem [{\citenamefont {Wildi}\ \emph {et~al.}(2019)\citenamefont {Wildi},
  \citenamefont {Brasch}, \citenamefont {Liu}, \citenamefont {Kippenberg},\
  and\ \citenamefont {Herr}}]{Wildi:19}%
  \BibitemOpen
  \bibfield  {author} {\bibinfo {author} {\bibfnamefont {T.}~\bibnamefont
  {Wildi}}, \bibinfo {author} {\bibfnamefont {V.}~\bibnamefont {Brasch}},
  \bibinfo {author} {\bibfnamefont {J.}~\bibnamefont {Liu}}, \bibinfo {author}
  {\bibfnamefont {T.~J.}\ \bibnamefont {Kippenberg}}, \ and\ \bibinfo {author}
  {\bibfnamefont {T.}~\bibnamefont {Herr}},\ }\bibfield  {title} {\enquote
  {\bibinfo {title} {Thermally stable access to microresonator solitons via
  slow pump modulation},}\ }\href {\doibase 10.1364/OL.44.004447} {\bibfield
  {journal} {\bibinfo  {journal} {Opt. Lett.}\ }\textbf {\bibinfo {volume}
  {44}},\ \bibinfo {pages} {4447} (\bibinfo {year} {2019})}\BibitemShut
  {NoStop}%
\bibitem [{\citenamefont {Nishimoto}\ \emph {et~al.}(2022)\citenamefont
  {Nishimoto}, \citenamefont {Minoshima}, \citenamefont {Yasui},\ and\
  \citenamefont {Kuse}}]{Nishimoto:22}%
  \BibitemOpen
  \bibfield  {author} {\bibinfo {author} {\bibfnamefont {K.}~\bibnamefont
  {Nishimoto}}, \bibinfo {author} {\bibfnamefont {K.}~\bibnamefont
  {Minoshima}}, \bibinfo {author} {\bibfnamefont {T.}~\bibnamefont {Yasui}}, \
  and\ \bibinfo {author} {\bibfnamefont {N.}~\bibnamefont {Kuse}},\ }\bibfield
  {title} {\enquote {\bibinfo {title} {Thermal control of a kerr microresonator
  soliton comb via an optical sideband},}\ }\href {\doibase 10.1364/OL.448326}
  {\bibfield  {journal} {\bibinfo  {journal} {Opt. Lett.}\ }\textbf {\bibinfo
  {volume} {47}},\ \bibinfo {pages} {281} (\bibinfo {year} {2022})}\BibitemShut
  {NoStop}%
\bibitem [{\citenamefont {Pavlov}\ \emph {et~al.}(2018)\citenamefont {Pavlov},
  \citenamefont {Koptyaev}, \citenamefont {Lihachev}, \citenamefont {Voloshin},
  \citenamefont {Gorodnitskiy}, \citenamefont {Ryabko}, \citenamefont
  {Polonsky},\ and\ \citenamefont {Gorodetsky}}]{Pavlov:18}%
  \BibitemOpen
  \bibfield  {author} {\bibinfo {author} {\bibfnamefont {N.~G.}\ \bibnamefont
  {Pavlov}}, \bibinfo {author} {\bibfnamefont {S.}~\bibnamefont {Koptyaev}},
  \bibinfo {author} {\bibfnamefont {G.~V.}\ \bibnamefont {Lihachev}}, \bibinfo
  {author} {\bibfnamefont {A.~S.}\ \bibnamefont {Voloshin}}, \bibinfo {author}
  {\bibfnamefont {A.~S.}\ \bibnamefont {Gorodnitskiy}}, \bibinfo {author}
  {\bibfnamefont {M.~V.}\ \bibnamefont {Ryabko}}, \bibinfo {author}
  {\bibfnamefont {S.~V.}\ \bibnamefont {Polonsky}}, \ and\ \bibinfo {author}
  {\bibfnamefont {M.~L.}\ \bibnamefont {Gorodetsky}},\ }\bibfield  {title}
  {\enquote {\bibinfo {title} {Narrow-linewidth lasing and soliton kerr
  microcombs with ordinary laser diodes},}\ }\href {\doibase
  10.1038/s41566-018-0277-2} {\bibfield  {journal} {\bibinfo  {journal} {Nature
  Photonics}\ }\textbf {\bibinfo {volume} {12}},\ \bibinfo {pages} {694}
  (\bibinfo {year} {2018})}\BibitemShut {NoStop}%
\bibitem [{\citenamefont {Kondratiev}\ \emph {et~al.}(2023)\citenamefont
  {Kondratiev}, \citenamefont {Lobanov}, \citenamefont {Shitikov},
  \citenamefont {Galiev}, \citenamefont {Chermoshentsev}, \citenamefont
  {Dmitriev}, \citenamefont {Danilin}, \citenamefont {Lonshakov}, \citenamefont
  {Min'kov}, \citenamefont {Sokol}, \citenamefont {Cordette}, \citenamefont
  {Luo}, \citenamefont {Liang}, \citenamefont {Liu},\ and\ \citenamefont
  {Bilenko}}]{Kondratiev:23}%
  \BibitemOpen
  \bibfield  {author} {\bibinfo {author} {\bibfnamefont {N.~M.}\ \bibnamefont
  {Kondratiev}}, \bibinfo {author} {\bibfnamefont {V.~E.}\ \bibnamefont
  {Lobanov}}, \bibinfo {author} {\bibfnamefont {A.~E.}\ \bibnamefont
  {Shitikov}}, \bibinfo {author} {\bibfnamefont {R.~R.}\ \bibnamefont
  {Galiev}}, \bibinfo {author} {\bibfnamefont {D.~A.}\ \bibnamefont
  {Chermoshentsev}}, \bibinfo {author} {\bibfnamefont {N.~Y.}\ \bibnamefont
  {Dmitriev}}, \bibinfo {author} {\bibfnamefont {A.~N.}\ \bibnamefont
  {Danilin}}, \bibinfo {author} {\bibfnamefont {E.~A.}\ \bibnamefont
  {Lonshakov}}, \bibinfo {author} {\bibfnamefont {K.~N.}\ \bibnamefont
  {Min'kov}}, \bibinfo {author} {\bibfnamefont {D.~M.}\ \bibnamefont {Sokol}},
  \bibinfo {author} {\bibfnamefont {S.~J.}\ \bibnamefont {Cordette}}, \bibinfo
  {author} {\bibfnamefont {Y.-H.}\ \bibnamefont {Luo}}, \bibinfo {author}
  {\bibfnamefont {W.}~\bibnamefont {Liang}}, \bibinfo {author} {\bibfnamefont
  {J.}~\bibnamefont {Liu}}, \ and\ \bibinfo {author} {\bibfnamefont {I.~A.}\
  \bibnamefont {Bilenko}},\ }\bibfield  {title} {\enquote {\bibinfo {title}
  {Recent advances in laser self-injection locking to high-q
  microresonators},}\ }\href {\doibase 10.1007/s11467-022-1245-3} {\bibfield
  {journal} {\bibinfo  {journal} {Frontiers of Physics}\ }\textbf {\bibinfo
  {volume} {18}},\ \bibinfo {pages} {21305} (\bibinfo {year}
  {2023})}\BibitemShut {NoStop}%
\bibitem [{\citenamefont {Shen}\ \emph {et~al.}(2020)\citenamefont {Shen},
  \citenamefont {Chang}, \citenamefont {Liu}, \citenamefont {Wang},
  \citenamefont {Yang}, \citenamefont {Xiang}, \citenamefont {Wang},
  \citenamefont {He}, \citenamefont {Liu}, \citenamefont {Xie}, \citenamefont
  {Guo}, \citenamefont {Kinghorn}, \citenamefont {Wu}, \citenamefont {Ji},
  \citenamefont {Kippenberg}, \citenamefont {Vahala},\ and\ \citenamefont
  {Bowers}}]{Shen:20}%
  \BibitemOpen
  \bibfield  {author} {\bibinfo {author} {\bibfnamefont {B.}~\bibnamefont
  {Shen}}, \bibinfo {author} {\bibfnamefont {L.}~\bibnamefont {Chang}},
  \bibinfo {author} {\bibfnamefont {J.}~\bibnamefont {Liu}}, \bibinfo {author}
  {\bibfnamefont {H.}~\bibnamefont {Wang}}, \bibinfo {author} {\bibfnamefont
  {Q.-F.}\ \bibnamefont {Yang}}, \bibinfo {author} {\bibfnamefont
  {C.}~\bibnamefont {Xiang}}, \bibinfo {author} {\bibfnamefont {R.~N.}\
  \bibnamefont {Wang}}, \bibinfo {author} {\bibfnamefont {J.}~\bibnamefont
  {He}}, \bibinfo {author} {\bibfnamefont {T.}~\bibnamefont {Liu}}, \bibinfo
  {author} {\bibfnamefont {W.}~\bibnamefont {Xie}}, \bibinfo {author}
  {\bibfnamefont {J.}~\bibnamefont {Guo}}, \bibinfo {author} {\bibfnamefont
  {D.}~\bibnamefont {Kinghorn}}, \bibinfo {author} {\bibfnamefont
  {L.}~\bibnamefont {Wu}}, \bibinfo {author} {\bibfnamefont {Q.-X.}\
  \bibnamefont {Ji}}, \bibinfo {author} {\bibfnamefont {T.~J.}\ \bibnamefont
  {Kippenberg}}, \bibinfo {author} {\bibfnamefont {K.}~\bibnamefont {Vahala}},
  \ and\ \bibinfo {author} {\bibfnamefont {J.~E.}\ \bibnamefont {Bowers}},\
  }\bibfield  {title} {\enquote {\bibinfo {title} {Integrated turnkey soliton
  microcombs},}\ }\href {\doibase 10.1038/s41586-020-2358-x} {\bibfield
  {journal} {\bibinfo  {journal} {Nature}\ }\textbf {\bibinfo {volume} {582}},\
  \bibinfo {pages} {365} (\bibinfo {year} {2020})}\BibitemShut {NoStop}%
\bibitem [{\citenamefont {Voloshin}\ \emph {et~al.}(2021)\citenamefont
  {Voloshin}, \citenamefont {Kondratiev}, \citenamefont {Lihachev},
  \citenamefont {Liu}, \citenamefont {Lobanov}, \citenamefont {Dmitriev},
  \citenamefont {Weng}, \citenamefont {Kippenberg},\ and\ \citenamefont
  {Bilenko}}]{Voloshin:21}%
  \BibitemOpen
  \bibfield  {author} {\bibinfo {author} {\bibfnamefont {A.~S.}\ \bibnamefont
  {Voloshin}}, \bibinfo {author} {\bibfnamefont {N.~M.}\ \bibnamefont
  {Kondratiev}}, \bibinfo {author} {\bibfnamefont {G.~V.}\ \bibnamefont
  {Lihachev}}, \bibinfo {author} {\bibfnamefont {J.}~\bibnamefont {Liu}},
  \bibinfo {author} {\bibfnamefont {V.~E.}\ \bibnamefont {Lobanov}}, \bibinfo
  {author} {\bibfnamefont {N.~Y.}\ \bibnamefont {Dmitriev}}, \bibinfo {author}
  {\bibfnamefont {W.}~\bibnamefont {Weng}}, \bibinfo {author} {\bibfnamefont
  {T.~J.}\ \bibnamefont {Kippenberg}}, \ and\ \bibinfo {author} {\bibfnamefont
  {I.~A.}\ \bibnamefont {Bilenko}},\ }\bibfield  {title} {\enquote {\bibinfo
  {title} {Dynamics of soliton self-injection locking in optical
  microresonators},}\ }\href {\doibase 10.1038/s41467-020-20196-y} {\bibfield
  {journal} {\bibinfo  {journal} {Nature Communications}\ }\textbf {\bibinfo
  {volume} {12}},\ \bibinfo {pages} {235} (\bibinfo {year} {2021})}\BibitemShut
  {NoStop}%
\bibitem [{\citenamefont {Briles}\ \emph {et~al.}(2021)\citenamefont {Briles},
  \citenamefont {Yu}, \citenamefont {Chang}, \citenamefont {Xiang},
  \citenamefont {Guo}, \citenamefont {Kinghorn}, \citenamefont {Moille},
  \citenamefont {Srinivasan}, \citenamefont {Bowers},\ and\ \citenamefont
  {Papp}}]{Briles:21}%
  \BibitemOpen
  \bibfield  {author} {\bibinfo {author} {\bibfnamefont {T.~C.}\ \bibnamefont
  {Briles}}, \bibinfo {author} {\bibfnamefont {S.-P.}\ \bibnamefont {Yu}},
  \bibinfo {author} {\bibfnamefont {L.}~\bibnamefont {Chang}}, \bibinfo
  {author} {\bibfnamefont {C.}~\bibnamefont {Xiang}}, \bibinfo {author}
  {\bibfnamefont {J.}~\bibnamefont {Guo}}, \bibinfo {author} {\bibfnamefont
  {D.}~\bibnamefont {Kinghorn}}, \bibinfo {author} {\bibfnamefont
  {G.}~\bibnamefont {Moille}}, \bibinfo {author} {\bibfnamefont
  {K.}~\bibnamefont {Srinivasan}}, \bibinfo {author} {\bibfnamefont {J.~E.}\
  \bibnamefont {Bowers}}, \ and\ \bibinfo {author} {\bibfnamefont {S.~B.}\
  \bibnamefont {Papp}},\ }\bibfield  {title} {\enquote {\bibinfo {title}
  {Hybrid inp and sin integration of an octave-spanning frequency comb},}\
  }\href {\doibase 10.1063/5.0035452} {\bibfield  {journal} {\bibinfo
  {journal} {APL Photonics}\ }\textbf {\bibinfo {volume} {6}},\ \bibinfo
  {pages} {026102} (\bibinfo {year} {2021})}\BibitemShut {NoStop}%
\bibitem [{\citenamefont {Savchenkov}\ \emph {et~al.}(2006)\citenamefont
  {Savchenkov}, \citenamefont {Matsko}, \citenamefont {Strekalov},
  \citenamefont {Ilchenko},\ and\ \citenamefont {Maleki}}]{Savchenkov:06}%
  \BibitemOpen
  \bibfield  {author} {\bibinfo {author} {\bibfnamefont {A.~A.}\ \bibnamefont
  {Savchenkov}}, \bibinfo {author} {\bibfnamefont {A.~B.}\ \bibnamefont
  {Matsko}}, \bibinfo {author} {\bibfnamefont {D.}~\bibnamefont {Strekalov}},
  \bibinfo {author} {\bibfnamefont {V.~S.}\ \bibnamefont {Ilchenko}}, \ and\
  \bibinfo {author} {\bibfnamefont {L.}~\bibnamefont {Maleki}},\ }\bibfield
  {title} {\enquote {\bibinfo {title} {Enhancement of photorefraction in
  whispering gallery mode resonators},}\ }\href {\doibase
  10.1103/PhysRevB.74.245119} {\bibfield  {journal} {\bibinfo  {journal} {Phys.
  Rev. B}\ }\textbf {\bibinfo {volume} {74}},\ \bibinfo {pages} {245119}
  (\bibinfo {year} {2006})}\BibitemShut {NoStop}%
\bibitem [{\citenamefont {Leidinger}\ \emph {et~al.}(2016)\citenamefont
  {Leidinger}, \citenamefont {Werner}, \citenamefont {Yoshiki}, \citenamefont
  {Buse},\ and\ \citenamefont {Breunig}}]{Leidinger:16}%
  \BibitemOpen
  \bibfield  {author} {\bibinfo {author} {\bibfnamefont {M.}~\bibnamefont
  {Leidinger}}, \bibinfo {author} {\bibfnamefont {C.~S.}\ \bibnamefont
  {Werner}}, \bibinfo {author} {\bibfnamefont {W.}~\bibnamefont {Yoshiki}},
  \bibinfo {author} {\bibfnamefont {K.}~\bibnamefont {Buse}}, \ and\ \bibinfo
  {author} {\bibfnamefont {I.}~\bibnamefont {Breunig}},\ }\bibfield  {title}
  {\enquote {\bibinfo {title} {Impact of the photorefractive and
  pyroelectric-electro-optic effect in lithium niobate on whispering-gallery
  modes},}\ }\href {\doibase 10.1364/OL.41.005474} {\bibfield  {journal}
  {\bibinfo  {journal} {Opt. Lett.}\ }\textbf {\bibinfo {volume} {41}},\
  \bibinfo {pages} {5474} (\bibinfo {year} {2016})}\BibitemShut {NoStop}%
\bibitem [{\citenamefont {Jiang}\ \emph {et~al.}(2017)\citenamefont {Jiang},
  \citenamefont {Luo}, \citenamefont {Liang}, \citenamefont {Chen},
  \citenamefont {Chen},\ and\ \citenamefont {Lin}}]{Jiang:17}%
  \BibitemOpen
  \bibfield  {author} {\bibinfo {author} {\bibfnamefont {H.}~\bibnamefont
  {Jiang}}, \bibinfo {author} {\bibfnamefont {R.}~\bibnamefont {Luo}}, \bibinfo
  {author} {\bibfnamefont {H.}~\bibnamefont {Liang}}, \bibinfo {author}
  {\bibfnamefont {X.}~\bibnamefont {Chen}}, \bibinfo {author} {\bibfnamefont
  {Y.}~\bibnamefont {Chen}}, \ and\ \bibinfo {author} {\bibfnamefont
  {Q.}~\bibnamefont {Lin}},\ }\bibfield  {title} {\enquote {\bibinfo {title}
  {Fast response of photorefraction in lithium niobate microresonators},}\
  }\href {\doibase 10.1364/OL.42.003267} {\bibfield  {journal} {\bibinfo
  {journal} {Opt. Lett.}\ }\textbf {\bibinfo {volume} {42}},\ \bibinfo {pages}
  {3267} (\bibinfo {year} {2017})}\BibitemShut {NoStop}%
\bibitem [{\citenamefont {Xu}\ \emph {et~al.}(2021{\natexlab{b}})\citenamefont
  {Xu}, \citenamefont {Shen}, \citenamefont {Lu}, \citenamefont {Surya},
  \citenamefont {Sayem},\ and\ \citenamefont {Tang}}]{XuY:21}%
  \BibitemOpen
  \bibfield  {author} {\bibinfo {author} {\bibfnamefont {Y.}~\bibnamefont
  {Xu}}, \bibinfo {author} {\bibfnamefont {M.}~\bibnamefont {Shen}}, \bibinfo
  {author} {\bibfnamefont {J.}~\bibnamefont {Lu}}, \bibinfo {author}
  {\bibfnamefont {J.~B.}\ \bibnamefont {Surya}}, \bibinfo {author}
  {\bibfnamefont {A.~A.}\ \bibnamefont {Sayem}}, \ and\ \bibinfo {author}
  {\bibfnamefont {H.~X.}\ \bibnamefont {Tang}},\ }\bibfield  {title} {\enquote
  {\bibinfo {title} {Mitigating photorefractive effect in thin-film lithium
  niobate microring resonators},}\ }\href {\doibase 10.1364/OE.418877}
  {\bibfield  {journal} {\bibinfo  {journal} {Opt. Express}\ }\textbf {\bibinfo
  {volume} {29}},\ \bibinfo {pages} {5497} (\bibinfo {year}
  {2021}{\natexlab{b}})}\BibitemShut {NoStop}%
\bibitem [{\citenamefont {Boes}\ \emph {et~al.}(2018)\citenamefont {Boes},
  \citenamefont {Corcoran}, \citenamefont {Chang}, \citenamefont {Bowers},\
  and\ \citenamefont {Mitchell}}]{Boes:18}%
  \BibitemOpen
  \bibfield  {author} {\bibinfo {author} {\bibfnamefont {A.}~\bibnamefont
  {Boes}}, \bibinfo {author} {\bibfnamefont {B.}~\bibnamefont {Corcoran}},
  \bibinfo {author} {\bibfnamefont {L.}~\bibnamefont {Chang}}, \bibinfo
  {author} {\bibfnamefont {J.}~\bibnamefont {Bowers}}, \ and\ \bibinfo {author}
  {\bibfnamefont {A.}~\bibnamefont {Mitchell}},\ }\bibfield  {title} {\enquote
  {\bibinfo {title} {Status and potential of lithium niobate on insulator
  (lnoi) for photonic integrated circuits},}\ }\href {\doibase
  https://doi.org/10.1002/lpor.201700256} {\bibfield  {journal} {\bibinfo
  {journal} {Laser \& Photonics Reviews}\ }\textbf {\bibinfo {volume} {12}},\
  \bibinfo {pages} {1700256} (\bibinfo {year} {2018})}\BibitemShut {NoStop}%
\bibitem [{\citenamefont {Zhu}\ \emph {et~al.}(2021)\citenamefont {Zhu},
  \citenamefont {Shao}, \citenamefont {Yu}, \citenamefont {Cheng},
  \citenamefont {Desiatov}, \citenamefont {Xin}, \citenamefont {Hu},
  \citenamefont {Holzgrafe}, \citenamefont {Ghosh}, \citenamefont
  {Shams-Ansari}, \citenamefont {Puma}, \citenamefont {Sinclair}, \citenamefont
  {Reimer}, \citenamefont {Zhang},\ and\ \citenamefont {Lon\v{c}ar}}]{Zhu:21}%
  \BibitemOpen
  \bibfield  {author} {\bibinfo {author} {\bibfnamefont {D.}~\bibnamefont
  {Zhu}}, \bibinfo {author} {\bibfnamefont {L.}~\bibnamefont {Shao}}, \bibinfo
  {author} {\bibfnamefont {M.}~\bibnamefont {Yu}}, \bibinfo {author}
  {\bibfnamefont {R.}~\bibnamefont {Cheng}}, \bibinfo {author} {\bibfnamefont
  {B.}~\bibnamefont {Desiatov}}, \bibinfo {author} {\bibfnamefont {C.~J.}\
  \bibnamefont {Xin}}, \bibinfo {author} {\bibfnamefont {Y.}~\bibnamefont
  {Hu}}, \bibinfo {author} {\bibfnamefont {J.}~\bibnamefont {Holzgrafe}},
  \bibinfo {author} {\bibfnamefont {S.}~\bibnamefont {Ghosh}}, \bibinfo
  {author} {\bibfnamefont {A.}~\bibnamefont {Shams-Ansari}}, \bibinfo {author}
  {\bibfnamefont {E.}~\bibnamefont {Puma}}, \bibinfo {author} {\bibfnamefont
  {N.}~\bibnamefont {Sinclair}}, \bibinfo {author} {\bibfnamefont
  {C.}~\bibnamefont {Reimer}}, \bibinfo {author} {\bibfnamefont
  {M.}~\bibnamefont {Zhang}}, \ and\ \bibinfo {author} {\bibfnamefont
  {M.}~\bibnamefont {Lon\v{c}ar}},\ }\bibfield  {title} {\enquote {\bibinfo
  {title} {Integrated photonics on thin-film lithium niobate},}\ }\href
  {\doibase 10.1364/AOP.411024} {\bibfield  {journal} {\bibinfo  {journal}
  {Adv. Opt. Photon.}\ }\textbf {\bibinfo {volume} {13}},\ \bibinfo {pages}
  {242} (\bibinfo {year} {2021})}\BibitemShut {NoStop}%
\bibitem [{\citenamefont {He}\ \emph {et~al.}(2019{\natexlab{a}})\citenamefont
  {He}, \citenamefont {Yang}, \citenamefont {Ling}, \citenamefont {Luo},
  \citenamefont {Liang}, \citenamefont {Li}, \citenamefont {Shen},
  \citenamefont {Wang}, \citenamefont {Vahala},\ and\ \citenamefont
  {Lin}}]{He:19}%
  \BibitemOpen
  \bibfield  {author} {\bibinfo {author} {\bibfnamefont {Y.}~\bibnamefont
  {He}}, \bibinfo {author} {\bibfnamefont {Q.-F.}\ \bibnamefont {Yang}},
  \bibinfo {author} {\bibfnamefont {J.}~\bibnamefont {Ling}}, \bibinfo {author}
  {\bibfnamefont {R.}~\bibnamefont {Luo}}, \bibinfo {author} {\bibfnamefont
  {H.}~\bibnamefont {Liang}}, \bibinfo {author} {\bibfnamefont
  {M.}~\bibnamefont {Li}}, \bibinfo {author} {\bibfnamefont {B.}~\bibnamefont
  {Shen}}, \bibinfo {author} {\bibfnamefont {H.}~\bibnamefont {Wang}}, \bibinfo
  {author} {\bibfnamefont {K.}~\bibnamefont {Vahala}}, \ and\ \bibinfo {author}
  {\bibfnamefont {Q.}~\bibnamefont {Lin}},\ }\bibfield  {title} {\enquote
  {\bibinfo {title} {Self-starting bi-chromatic linbo3 soliton microcomb},}\
  }\href {\doibase 10.1364/OPTICA.6.001138} {\bibfield  {journal} {\bibinfo
  {journal} {Optica}\ }\textbf {\bibinfo {volume} {6}},\ \bibinfo {pages}
  {1138} (\bibinfo {year} {2019}{\natexlab{a}})}\BibitemShut {NoStop}%
\bibitem [{\citenamefont {Bruch}\ \emph {et~al.}(2021)\citenamefont {Bruch},
  \citenamefont {Liu}, \citenamefont {Gong}, \citenamefont {Surya},
  \citenamefont {Li}, \citenamefont {Zou},\ and\ \citenamefont
  {Tang}}]{Bruch:21}%
  \BibitemOpen
  \bibfield  {author} {\bibinfo {author} {\bibfnamefont {A.~W.}\ \bibnamefont
  {Bruch}}, \bibinfo {author} {\bibfnamefont {X.}~\bibnamefont {Liu}}, \bibinfo
  {author} {\bibfnamefont {Z.}~\bibnamefont {Gong}}, \bibinfo {author}
  {\bibfnamefont {J.~B.}\ \bibnamefont {Surya}}, \bibinfo {author}
  {\bibfnamefont {M.}~\bibnamefont {Li}}, \bibinfo {author} {\bibfnamefont
  {C.-L.}\ \bibnamefont {Zou}}, \ and\ \bibinfo {author} {\bibfnamefont
  {H.~X.}\ \bibnamefont {Tang}},\ }\bibfield  {title} {\enquote {\bibinfo
  {title} {Pockels soliton microcomb},}\ }\href {\doibase
  10.1038/s41566-020-00704-8} {\bibfield  {journal} {\bibinfo  {journal}
  {Nature Photonics}\ }\textbf {\bibinfo {volume} {15}},\ \bibinfo {pages} {21}
  (\bibinfo {year} {2021})}\BibitemShut {NoStop}%
\bibitem [{\citenamefont {He}\ \emph {et~al.}(2019{\natexlab{b}})\citenamefont
  {He}, \citenamefont {Xu}, \citenamefont {Ren}, \citenamefont {Jian},
  \citenamefont {Ruan}, \citenamefont {Xu}, \citenamefont {Gao}, \citenamefont
  {Sun}, \citenamefont {Wen}, \citenamefont {Zhou}, \citenamefont {Liu},
  \citenamefont {Guo}, \citenamefont {Chen}, \citenamefont {Yu}, \citenamefont
  {Liu},\ and\ \citenamefont {Cai}}]{HeM:19}%
  \BibitemOpen
  \bibfield  {author} {\bibinfo {author} {\bibfnamefont {M.}~\bibnamefont
  {He}}, \bibinfo {author} {\bibfnamefont {M.}~\bibnamefont {Xu}}, \bibinfo
  {author} {\bibfnamefont {Y.}~\bibnamefont {Ren}}, \bibinfo {author}
  {\bibfnamefont {J.}~\bibnamefont {Jian}}, \bibinfo {author} {\bibfnamefont
  {Z.}~\bibnamefont {Ruan}}, \bibinfo {author} {\bibfnamefont {Y.}~\bibnamefont
  {Xu}}, \bibinfo {author} {\bibfnamefont {S.}~\bibnamefont {Gao}}, \bibinfo
  {author} {\bibfnamefont {S.}~\bibnamefont {Sun}}, \bibinfo {author}
  {\bibfnamefont {X.}~\bibnamefont {Wen}}, \bibinfo {author} {\bibfnamefont
  {L.}~\bibnamefont {Zhou}}, \bibinfo {author} {\bibfnamefont {L.}~\bibnamefont
  {Liu}}, \bibinfo {author} {\bibfnamefont {C.}~\bibnamefont {Guo}}, \bibinfo
  {author} {\bibfnamefont {H.}~\bibnamefont {Chen}}, \bibinfo {author}
  {\bibfnamefont {S.}~\bibnamefont {Yu}}, \bibinfo {author} {\bibfnamefont
  {L.}~\bibnamefont {Liu}}, \ and\ \bibinfo {author} {\bibfnamefont
  {X.}~\bibnamefont {Cai}},\ }\bibfield  {title} {\enquote {\bibinfo {title}
  {High-performance hybrid silicon and lithium niobate mach--zehnder modulators
  for 100 gbit s-1 and beyond},}\ }\href {\doibase 10.1038/s41566-019-0378-6}
  {\bibfield  {journal} {\bibinfo  {journal} {Nature Photonics}\ }\textbf
  {\bibinfo {volume} {13}},\ \bibinfo {pages} {359} (\bibinfo {year}
  {2019}{\natexlab{b}})}\BibitemShut {NoStop}%
\bibitem [{\citenamefont {Wang}\ \emph {et~al.}(2018)\citenamefont {Wang},
  \citenamefont {Zhang}, \citenamefont {Chen}, \citenamefont {Bertrand},
  \citenamefont {Shams-Ansari}, \citenamefont {Chandrasekhar}, \citenamefont
  {Winzer},\ and\ \citenamefont {Lon{\v c}ar}}]{WangC:18}%
  \BibitemOpen
  \bibfield  {author} {\bibinfo {author} {\bibfnamefont {C.}~\bibnamefont
  {Wang}}, \bibinfo {author} {\bibfnamefont {M.}~\bibnamefont {Zhang}},
  \bibinfo {author} {\bibfnamefont {X.}~\bibnamefont {Chen}}, \bibinfo {author}
  {\bibfnamefont {M.}~\bibnamefont {Bertrand}}, \bibinfo {author}
  {\bibfnamefont {A.}~\bibnamefont {Shams-Ansari}}, \bibinfo {author}
  {\bibfnamefont {S.}~\bibnamefont {Chandrasekhar}}, \bibinfo {author}
  {\bibfnamefont {P.}~\bibnamefont {Winzer}}, \ and\ \bibinfo {author}
  {\bibfnamefont {M.}~\bibnamefont {Lon{\v c}ar}},\ }\bibfield  {title}
  {\enquote {\bibinfo {title} {Integrated lithium niobate electro-optic
  modulators operating at cmos-compatible voltages},}\ }\href {\doibase
  10.1038/s41586-018-0551-y} {\bibfield  {journal} {\bibinfo  {journal}
  {Nature}\ }\textbf {\bibinfo {volume} {562}},\ \bibinfo {pages} {101}
  (\bibinfo {year} {2018})}\BibitemShut {NoStop}%
\bibitem [{\citenamefont {Wang}\ \emph {et~al.}(2019)\citenamefont {Wang},
  \citenamefont {Zhang}, \citenamefont {Yu}, \citenamefont {Zhu}, \citenamefont
  {Hu},\ and\ \citenamefont {Loncar}}]{Wang:19}%
  \BibitemOpen
  \bibfield  {author} {\bibinfo {author} {\bibfnamefont {C.}~\bibnamefont
  {Wang}}, \bibinfo {author} {\bibfnamefont {M.}~\bibnamefont {Zhang}},
  \bibinfo {author} {\bibfnamefont {M.}~\bibnamefont {Yu}}, \bibinfo {author}
  {\bibfnamefont {R.}~\bibnamefont {Zhu}}, \bibinfo {author} {\bibfnamefont
  {H.}~\bibnamefont {Hu}}, \ and\ \bibinfo {author} {\bibfnamefont
  {M.}~\bibnamefont {Loncar}},\ }\bibfield  {title} {\enquote {\bibinfo {title}
  {Monolithic lithium niobate photonic circuits for kerr frequency comb
  generation and modulation},}\ }\href {\doibase 10.1038/s41467-019-08969-6}
  {\bibfield  {journal} {\bibinfo  {journal} {Nature Communications}\ }\textbf
  {\bibinfo {volume} {10}},\ \bibinfo {pages} {978} (\bibinfo {year}
  {2019})}\BibitemShut {NoStop}%
\bibitem [{\citenamefont {Fang}\ \emph {et~al.}(2019)\citenamefont {Fang},
  \citenamefont {Luo}, \citenamefont {Lin}, \citenamefont {Wang}, \citenamefont
  {Zhang}, \citenamefont {Wu}, \citenamefont {Zhou}, \citenamefont {Chu},
  \citenamefont {Lu},\ and\ \citenamefont {Cheng}}]{Fang:19}%
  \BibitemOpen
  \bibfield  {author} {\bibinfo {author} {\bibfnamefont {Z.}~\bibnamefont
  {Fang}}, \bibinfo {author} {\bibfnamefont {H.}~\bibnamefont {Luo}}, \bibinfo
  {author} {\bibfnamefont {J.}~\bibnamefont {Lin}}, \bibinfo {author}
  {\bibfnamefont {M.}~\bibnamefont {Wang}}, \bibinfo {author} {\bibfnamefont
  {J.}~\bibnamefont {Zhang}}, \bibinfo {author} {\bibfnamefont
  {R.}~\bibnamefont {Wu}}, \bibinfo {author} {\bibfnamefont {J.}~\bibnamefont
  {Zhou}}, \bibinfo {author} {\bibfnamefont {W.}~\bibnamefont {Chu}}, \bibinfo
  {author} {\bibfnamefont {T.}~\bibnamefont {Lu}}, \ and\ \bibinfo {author}
  {\bibfnamefont {Y.}~\bibnamefont {Cheng}},\ }\bibfield  {title} {\enquote
  {\bibinfo {title} {Efficient electro-optical tuning of an optical frequency
  microcomb on a monolithically integrated high-q lithium niobate microdisk},}\
  }\href {\doibase 10.1364/OL.44.005953} {\bibfield  {journal} {\bibinfo
  {journal} {Opt. Lett.}\ }\textbf {\bibinfo {volume} {44}},\ \bibinfo {pages}
  {5953} (\bibinfo {year} {2019})}\BibitemShut {NoStop}%
\bibitem [{\citenamefont {Gorodetsky}\ \emph {et~al.}(2000)\citenamefont
  {Gorodetsky}, \citenamefont {Pryamikov},\ and\ \citenamefont
  {Ilchenko}}]{Gorodetsky:00}%
  \BibitemOpen
  \bibfield  {author} {\bibinfo {author} {\bibfnamefont {M.~L.}\ \bibnamefont
  {Gorodetsky}}, \bibinfo {author} {\bibfnamefont {A.~D.}\ \bibnamefont
  {Pryamikov}}, \ and\ \bibinfo {author} {\bibfnamefont {V.~S.}\ \bibnamefont
  {Ilchenko}},\ }\bibfield  {title} {\enquote {\bibinfo {title} {Rayleigh
  scattering in high-q microspheres},}\ }\href {\doibase
  10.1364/JOSAB.17.001051} {\bibfield  {journal} {\bibinfo  {journal} {J. Opt.
  Soc. Am. B}\ }\textbf {\bibinfo {volume} {17}},\ \bibinfo {pages} {1051}
  (\bibinfo {year} {2000})}\BibitemShut {NoStop}%
\bibitem [{\citenamefont {Mili\'{a}n}\ and\ \citenamefont
  {Skryabin}(2014)}]{Milian:14}%
  \BibitemOpen
  \bibfield  {author} {\bibinfo {author} {\bibfnamefont {C.}~\bibnamefont
  {Mili\'{a}n}}\ and\ \bibinfo {author} {\bibfnamefont {D.}~\bibnamefont
  {Skryabin}},\ }\bibfield  {title} {\enquote {\bibinfo {title} {Soliton
  families and resonant radiation in a micro-ring resonator near zero
  group-velocity dispersion},}\ }\href {\doibase 10.1364/OE.22.003732}
  {\bibfield  {journal} {\bibinfo  {journal} {Opt. Express}\ }\textbf {\bibinfo
  {volume} {22}},\ \bibinfo {pages} {3732} (\bibinfo {year}
  {2014})}\BibitemShut {NoStop}%
\bibitem [{\citenamefont {Jang}\ \emph {et~al.}(2014)\citenamefont {Jang},
  \citenamefont {Erkintalo}, \citenamefont {Murdoch},\ and\ \citenamefont
  {Coen}}]{Jang:14}%
  \BibitemOpen
  \bibfield  {author} {\bibinfo {author} {\bibfnamefont {J.~K.}\ \bibnamefont
  {Jang}}, \bibinfo {author} {\bibfnamefont {M.}~\bibnamefont {Erkintalo}},
  \bibinfo {author} {\bibfnamefont {S.~G.}\ \bibnamefont {Murdoch}}, \ and\
  \bibinfo {author} {\bibfnamefont {S.}~\bibnamefont {Coen}},\ }\bibfield
  {title} {\enquote {\bibinfo {title} {Observation of dispersive wave emission
  by temporal cavity solitons},}\ }\href {\doibase 10.1364/OL.39.005503}
  {\bibfield  {journal} {\bibinfo  {journal} {Opt. Lett.}\ }\textbf {\bibinfo
  {volume} {39}},\ \bibinfo {pages} {5503} (\bibinfo {year}
  {2014})}\BibitemShut {NoStop}%
\bibitem [{\citenamefont {Yu}\ \emph {et~al.}(2020)\citenamefont {Yu},
  \citenamefont {Okawachi}, \citenamefont {Cheng}, \citenamefont {Wang},
  \citenamefont {Zhang}, \citenamefont {Gaeta},\ and\ \citenamefont {Lon{\v
  c}ar}}]{YuM:20}%
  \BibitemOpen
  \bibfield  {author} {\bibinfo {author} {\bibfnamefont {M.}~\bibnamefont
  {Yu}}, \bibinfo {author} {\bibfnamefont {Y.}~\bibnamefont {Okawachi}},
  \bibinfo {author} {\bibfnamefont {R.}~\bibnamefont {Cheng}}, \bibinfo
  {author} {\bibfnamefont {C.}~\bibnamefont {Wang}}, \bibinfo {author}
  {\bibfnamefont {M.}~\bibnamefont {Zhang}}, \bibinfo {author} {\bibfnamefont
  {A.~L.}\ \bibnamefont {Gaeta}}, \ and\ \bibinfo {author} {\bibfnamefont
  {M.}~\bibnamefont {Lon{\v c}ar}},\ }\bibfield  {title} {\enquote {\bibinfo
  {title} {Raman lasing and soliton mode-locking in lithium niobate
  microresonators},}\ }\href {\doibase 10.1038/s41377-020-0246-7} {\bibfield
  {journal} {\bibinfo  {journal} {Light: Science \& Applications}\ }\textbf
  {\bibinfo {volume} {9}},\ \bibinfo {pages} {9} (\bibinfo {year}
  {2020})}\BibitemShut {NoStop}%
\bibitem [{\citenamefont {Niu}\ \emph {et~al.}(2023)\citenamefont {Niu},
  \citenamefont {Li}, \citenamefont {Wan}, \citenamefont {Sun}, \citenamefont
  {Hu}, \citenamefont {Zou}, \citenamefont {Guo},\ and\ \citenamefont
  {Dong}}]{Niu:23}%
  \BibitemOpen
  \bibfield  {author} {\bibinfo {author} {\bibfnamefont {R.}~\bibnamefont
  {Niu}}, \bibinfo {author} {\bibfnamefont {M.}~\bibnamefont {Li}}, \bibinfo
  {author} {\bibfnamefont {S.}~\bibnamefont {Wan}}, \bibinfo {author}
  {\bibfnamefont {Y.~R.}\ \bibnamefont {Sun}}, \bibinfo {author} {\bibfnamefont
  {S.-M.}\ \bibnamefont {Hu}}, \bibinfo {author} {\bibfnamefont {C.-L.}\
  \bibnamefont {Zou}}, \bibinfo {author} {\bibfnamefont {G.-C.}\ \bibnamefont
  {Guo}}, \ and\ \bibinfo {author} {\bibfnamefont {C.-H.}\ \bibnamefont
  {Dong}},\ }\bibfield  {title} {\enquote {\bibinfo {title} {khz-precision
  wavemeter based on reconfigurable microsoliton},}\ }\href {\doibase
  10.1038/s41467-022-35728-x} {\bibfield  {journal} {\bibinfo  {journal}
  {Nature Communications}\ }\textbf {\bibinfo {volume} {14}},\ \bibinfo {pages}
  {169} (\bibinfo {year} {2023})}\BibitemShut {NoStop}%
\bibitem [{\citenamefont {Li}\ \emph {et~al.}(2022{\natexlab{a}})\citenamefont
  {Li}, \citenamefont {Wan}, \citenamefont {Peng}, \citenamefont {Wang},
  \citenamefont {Niu}, \citenamefont {Zou}, \citenamefont {Guo},\ and\
  \citenamefont {Dong}}]{LiJ:22}%
  \BibitemOpen
  \bibfield  {author} {\bibinfo {author} {\bibfnamefont {J.}~\bibnamefont
  {Li}}, \bibinfo {author} {\bibfnamefont {S.}~\bibnamefont {Wan}}, \bibinfo
  {author} {\bibfnamefont {J.-L.}\ \bibnamefont {Peng}}, \bibinfo {author}
  {\bibfnamefont {Z.-Y.}\ \bibnamefont {Wang}}, \bibinfo {author}
  {\bibfnamefont {R.}~\bibnamefont {Niu}}, \bibinfo {author} {\bibfnamefont
  {C.-L.}\ \bibnamefont {Zou}}, \bibinfo {author} {\bibfnamefont {G.-C.}\
  \bibnamefont {Guo}}, \ and\ \bibinfo {author} {\bibfnamefont {C.-H.}\
  \bibnamefont {Dong}},\ }\bibfield  {title} {\enquote {\bibinfo {title}
  {Thermal tuning of mode crossing and the perfect soliton crystal in a si3n4
  microresonator},}\ }\href {\doibase 10.1364/OE.450100} {\bibfield  {journal}
  {\bibinfo  {journal} {Opt. Express}\ }\textbf {\bibinfo {volume} {30}},\
  \bibinfo {pages} {13690} (\bibinfo {year} {2022}{\natexlab{a}})}\BibitemShut
  {NoStop}%
\bibitem [{\citenamefont {Guo}\ \emph {et~al.}(2017)\citenamefont {Guo},
  \citenamefont {Karpov}, \citenamefont {Lucas}, \citenamefont {Kordts},
  \citenamefont {Pfeiffer}, \citenamefont {Brasch}, \citenamefont {Lihachev},
  \citenamefont {Lobanov}, \citenamefont {Gorodetsky},\ and\ \citenamefont
  {Kippenberg}}]{Guo:16}%
  \BibitemOpen
  \bibfield  {author} {\bibinfo {author} {\bibfnamefont {H.}~\bibnamefont
  {Guo}}, \bibinfo {author} {\bibfnamefont {M.}~\bibnamefont {Karpov}},
  \bibinfo {author} {\bibfnamefont {E.}~\bibnamefont {Lucas}}, \bibinfo
  {author} {\bibfnamefont {A.}~\bibnamefont {Kordts}}, \bibinfo {author}
  {\bibfnamefont {M.~H.~P.}\ \bibnamefont {Pfeiffer}}, \bibinfo {author}
  {\bibfnamefont {V.}~\bibnamefont {Brasch}}, \bibinfo {author} {\bibfnamefont
  {G.}~\bibnamefont {Lihachev}}, \bibinfo {author} {\bibfnamefont {V.~E.}\
  \bibnamefont {Lobanov}}, \bibinfo {author} {\bibfnamefont {M.~L.}\
  \bibnamefont {Gorodetsky}}, \ and\ \bibinfo {author} {\bibfnamefont {T.~J.}\
  \bibnamefont {Kippenberg}},\ }\bibfield  {title} {\enquote {\bibinfo {title}
  {Universal dynamics and deterministic switching of dissipative kerr solitons
  in optical microresonators},}\ }\href {\doibase 10.1038/nphys3893} {\bibfield
   {journal} {\bibinfo  {journal} {Nature Physics}\ }\textbf {\bibinfo {volume}
  {13}},\ \bibinfo {pages} {94} (\bibinfo {year} {2017})}\BibitemShut {NoStop}%
\bibitem [{\citenamefont {Shams-Ansari}\ \emph {et~al.}(2022)\citenamefont
  {Shams-Ansari}, \citenamefont {Renaud}, \citenamefont {Cheng}, \citenamefont
  {Shao}, \citenamefont {He}, \citenamefont {Zhu}, \citenamefont {Yu},
  \citenamefont {Grant}, \citenamefont {Johansson}, \citenamefont {Zhang},\
  and\ \citenamefont {Lon\v{c}ar}}]{Shams-Ansari:22}%
  \BibitemOpen
  \bibfield  {author} {\bibinfo {author} {\bibfnamefont {A.}~\bibnamefont
  {Shams-Ansari}}, \bibinfo {author} {\bibfnamefont {D.}~\bibnamefont
  {Renaud}}, \bibinfo {author} {\bibfnamefont {R.}~\bibnamefont {Cheng}},
  \bibinfo {author} {\bibfnamefont {L.}~\bibnamefont {Shao}}, \bibinfo {author}
  {\bibfnamefont {L.}~\bibnamefont {He}}, \bibinfo {author} {\bibfnamefont
  {D.}~\bibnamefont {Zhu}}, \bibinfo {author} {\bibfnamefont {M.}~\bibnamefont
  {Yu}}, \bibinfo {author} {\bibfnamefont {H.~R.}\ \bibnamefont {Grant}},
  \bibinfo {author} {\bibfnamefont {L.}~\bibnamefont {Johansson}}, \bibinfo
  {author} {\bibfnamefont {M.}~\bibnamefont {Zhang}}, \ and\ \bibinfo {author}
  {\bibfnamefont {M.}~\bibnamefont {Lon\v{c}ar}},\ }\bibfield  {title}
  {\enquote {\bibinfo {title} {Electrically pumped laser transmitter integrated
  on thin-film lithium niobate},}\ }\href {\doibase 10.1364/OPTICA.448617}
  {\bibfield  {journal} {\bibinfo  {journal} {Optica}\ }\textbf {\bibinfo
  {volume} {9}},\ \bibinfo {pages} {408} (\bibinfo {year} {2022})}\BibitemShut
  {NoStop}%
\bibitem [{\citenamefont {Li}\ \emph {et~al.}(2022{\natexlab{b}})\citenamefont
  {Li}, \citenamefont {Chang}, \citenamefont {Wu}, \citenamefont {Staffa},
  \citenamefont {Ling}, \citenamefont {Javid}, \citenamefont {Xue},
  \citenamefont {He}, \citenamefont {Lopez-rios}, \citenamefont {Morin},
  \citenamefont {Wang}, \citenamefont {Shen}, \citenamefont {Zeng},
  \citenamefont {Zhu}, \citenamefont {Vahala}, \citenamefont {Bowers},\ and\
  \citenamefont {Lin}}]{LiM:22}%
  \BibitemOpen
  \bibfield  {author} {\bibinfo {author} {\bibfnamefont {M.}~\bibnamefont
  {Li}}, \bibinfo {author} {\bibfnamefont {L.}~\bibnamefont {Chang}}, \bibinfo
  {author} {\bibfnamefont {L.}~\bibnamefont {Wu}}, \bibinfo {author}
  {\bibfnamefont {J.}~\bibnamefont {Staffa}}, \bibinfo {author} {\bibfnamefont
  {J.}~\bibnamefont {Ling}}, \bibinfo {author} {\bibfnamefont {U.~A.}\
  \bibnamefont {Javid}}, \bibinfo {author} {\bibfnamefont {S.}~\bibnamefont
  {Xue}}, \bibinfo {author} {\bibfnamefont {Y.}~\bibnamefont {He}}, \bibinfo
  {author} {\bibfnamefont {R.}~\bibnamefont {Lopez-rios}}, \bibinfo {author}
  {\bibfnamefont {T.~J.}\ \bibnamefont {Morin}}, \bibinfo {author}
  {\bibfnamefont {H.}~\bibnamefont {Wang}}, \bibinfo {author} {\bibfnamefont
  {B.}~\bibnamefont {Shen}}, \bibinfo {author} {\bibfnamefont {S.}~\bibnamefont
  {Zeng}}, \bibinfo {author} {\bibfnamefont {L.}~\bibnamefont {Zhu}}, \bibinfo
  {author} {\bibfnamefont {K.~J.}\ \bibnamefont {Vahala}}, \bibinfo {author}
  {\bibfnamefont {J.~E.}\ \bibnamefont {Bowers}}, \ and\ \bibinfo {author}
  {\bibfnamefont {Q.}~\bibnamefont {Lin}},\ }\bibfield  {title} {\enquote
  {\bibinfo {title} {Integrated pockels laser},}\ }\href {\doibase
  10.1038/s41467-022-33101-6} {\bibfield  {journal} {\bibinfo  {journal}
  {Nature Communications}\ }\textbf {\bibinfo {volume} {13}},\ \bibinfo {pages}
  {5344} (\bibinfo {year} {2022}{\natexlab{b}})}\BibitemShut {NoStop}%
\bibitem [{\citenamefont {Snigirev}\ \emph {et~al.}(2023)\citenamefont
  {Snigirev}, \citenamefont {Riedhauser}, \citenamefont {Lihachev},
  \citenamefont {Churaev}, \citenamefont {Riemensberger}, \citenamefont {Wang},
  \citenamefont {Siddharth}, \citenamefont {Huang}, \citenamefont {M{\"o}hl},
  \citenamefont {Popoff}, \citenamefont {Drechsler}, \citenamefont {Caimi},
  \citenamefont {H{\"o}nl}, \citenamefont {Liu}, \citenamefont {Seidler},\ and\
  \citenamefont {Kippenberg}}]{Snigirev:23}%
  \BibitemOpen
  \bibfield  {author} {\bibinfo {author} {\bibfnamefont {V.}~\bibnamefont
  {Snigirev}}, \bibinfo {author} {\bibfnamefont {A.}~\bibnamefont
  {Riedhauser}}, \bibinfo {author} {\bibfnamefont {G.}~\bibnamefont
  {Lihachev}}, \bibinfo {author} {\bibfnamefont {M.}~\bibnamefont {Churaev}},
  \bibinfo {author} {\bibfnamefont {J.}~\bibnamefont {Riemensberger}}, \bibinfo
  {author} {\bibfnamefont {R.~N.}\ \bibnamefont {Wang}}, \bibinfo {author}
  {\bibfnamefont {A.}~\bibnamefont {Siddharth}}, \bibinfo {author}
  {\bibfnamefont {G.}~\bibnamefont {Huang}}, \bibinfo {author} {\bibfnamefont
  {C.}~\bibnamefont {M{\"o}hl}}, \bibinfo {author} {\bibfnamefont
  {Y.}~\bibnamefont {Popoff}}, \bibinfo {author} {\bibfnamefont
  {U.}~\bibnamefont {Drechsler}}, \bibinfo {author} {\bibfnamefont
  {D.}~\bibnamefont {Caimi}}, \bibinfo {author} {\bibfnamefont
  {S.}~\bibnamefont {H{\"o}nl}}, \bibinfo {author} {\bibfnamefont
  {J.}~\bibnamefont {Liu}}, \bibinfo {author} {\bibfnamefont {P.}~\bibnamefont
  {Seidler}}, \ and\ \bibinfo {author} {\bibfnamefont {T.~J.}\ \bibnamefont
  {Kippenberg}},\ }\bibfield  {title} {\enquote {\bibinfo {title} {Ultrafast
  tunable lasers using lithium niobate integrated photonics},}\ }\href
  {\doibase 10.1038/s41586-023-05724-2} {\bibfield  {journal} {\bibinfo
  {journal} {Nature}\ }\textbf {\bibinfo {volume} {615}},\ \bibinfo {pages}
  {411} (\bibinfo {year} {2023})}\BibitemShut {NoStop}%
\end{thebibliography}
%

\vbox{}

\noindent\textbf{Acknowledgments}\\
The work was supported by the Innovation program for Quantum Science and Technology (2021ZD0303203),
the National Natural Science Foundation of China (12293052, 11934012,12104442, 92050109, and 92250302),
the CAS Project for Young Scientists in Basic Research (YSBR-069),
the Fundamental Research Funds for the Central Universities.
This work was partially carried out at the USTC Center for Micro and Nanoscale Research and Fabrication.

\vbox{}

\noindent\textbf{Author contributions}\\
S.W., and P.-Y.W. contribute equally to this work.
S.W., P.-Y.W. and C.-H. D. conceived the experiments.
S.W., P. -Y. W., R.M. and F. B. prepared devices.
S.W., and P.-Y.W.built the experimental setup and performed experiments, with assistance
from Z.-Y. W., R. N., D. Y. H. and J. L..
S.W., J. L. and C.-H.D. wrote the manuscript with input from all co-authors.
C.-H.D. and G.-C.G. supervised the project.
All authors contributed extensively to the work presented in this paper.

\vbox{}

\noindent\textbf{Competing financial interests}\\
The authors declare
no competing financial interests.

\vbox{}

\noindent\textbf{Additional information}\\
Supplementary information is available in the online version of the paper.
Correspondence and requests for materials should be addressed to Fang Bo or Chun-Hua Dong.

\end{document}